\documentclass{ws-ijmpd}
\usepackage[super,compress]{cite}
\begin{document}

\markboth{Jayanne English}
{Cosmos and Canvas}

%
\catchline{}{}{}{}{}
%

\title{Canvas and Cosmos: Visual Art Techniques Applied to Astronomy Data.}

\author{JAYANNE ENGLISH\footnote{Permanent Address}}

\address{Department of Physics and Astronomy, University of Manitoba, \\
Winnipeg, Manitoba, R3T 2N2,
Canada.\\
Jayanne\_English@umanitoba.ca}

\maketitle

\begin{history}
\received{Day Month Year}
\revised{Day Month Year}
\end{history}

\begin{abstract}

Bold colour images from telescopes act as extraordinary ambassadors for research astronomers because they pique the public's curiosity.  But are they snapshots documenting physical reality?  Or are we looking at artistic spacescapes created by digitally manipulating astronomy images? This paper provides a tour of how original black and white data, from all regimes of the electromagnetic spectrum, are converted into the colour images gracing popular magazines, numerous websites, and even clothing. The history and method of the technical construction of these images is outlined. However,  the paper focuses on introducing the scientific reader to visual literacy (e.\/g.~human perception)  and  techniques from art (e.\/g.~composition, colour theory) since  these techniques can produce not only striking but politically powerful public outreach images. When created by research astronomers, the cultures of science and visual art can be balanced and the image can illuminate scientific results sufficiently strongly that the images are also used in research publications. Included are reflections on how they could feedback into astronomy research endeavours and future forms of visualization as well as  on the relevance of outreach images to visual art.  {\it (See the colour online version, in which figures can be enlarged, at http://xxxxxxx.) }
\end{abstract}


\ccode{PACS numbers:}


\section{Introduction: Professional Astronomers and Public Images} 	
\label{intro}\label{intropower}

Currently in astronomy it is not unusual for compellingly realistic, colourful images  to appear in professional  journals. Many are pictures that were originally created, by astronomers and out of scientific data, to promote professional astrophysical research endeavours to the public. These public outreach images are ubiquitous, appearing not only in traditional media such as textbooks, newspapers and magazines, but also on virtual telescope websites\footnote{E.\/g.~WorldWide Telescope, Google Sky.}, astronomy image galleries\footnote{E.\/g.~From the Earth to the Universe, NASA's Astronomy Picture of the Day (APOD) and HubbleSite, CFHT Astronomy Image of the Month, ASTRON JIVE Daily Image, ATNF Daily Astronomy Image.), online blogs, video podcasts  (e.\/g.~Sky at Night Magazine Vodcast, Hubblecast},  television\footnote{E.\/g/~Star Trek, The Big Bang Theory, as well as astronomy specific shows such as Cosmos.}, etc.   Besides fulfilling their conventional aim of encouraging the public to attend lectures and planetaria, these images are becoming cultural icons,  appearing in unexpected places such as fashionable garments\footnote{E.\/g.~Christopher Kane 2011 Resort Collection; ShadowPlay NYC.}, accessories\footnote{E.\/g.~Galaxy Backpack from ThinkGeek.} and designer home furnishings\footnote{E.\/g.~Shoenstaub Nebula Collection area rugs; Cosmic Mugs from Cherrico pottery.}.    In terms of political influence, astronomy images certainly helped fuel the public's outcry against the 2004 announcement to cancel service of the Hubble Space Telescope (HST), helping cause the reversal of this unpopular decision.  The appropriation of these striking outreach images into research papers was unanticipated and is another demonstration of their positive impact. 

Recognizing the power of outreach activities  the International Astronomical Union issued the following 2005 statement to  {\it professional} researchers, encouraging them to gear up for the International Year of Astronomy in 2009:

``As our world grows ever more complex and the pace of scientific discovery and technological change quickens, the global community of professional astronomers needs to communicate more effectively with the public.''\cite{iaustatement} 

In this paper I will therefore focus on the public images created by professional astronomers, although citizen scientists (i.\/e.~amateur astronomers) are strong contributors to this field,  often providing perspective on the public's astronomy interest to the research focused image-makers.  However, unlike the images made by amateur astronomers, those made by research astronomers are highly informed by their scientific measurements and are strongly affected by scientific culture, which includes a gravitation towards conventional scientific visual practices (e.\/g.~comprehending data through the use of contour plots such as the one in \S~\ref{futureastro}). 

Providing a well-known example of professional engagement  is the Hubble Heritage Team, which creates striking HST images for monthly release on the Hubble Heritage website at the Space Telescope Science Institute. 
During its 1998 inception the team,  consisting of 1 amateur astronomer and 7 people with post-graduate degrees in astronomy,   articulated the goals of their public outreach images as follows: 

``By emphasizing compelling HST images distilled from scientific data, we hope to pique curiosity about our astrophysical understanding of the universe we all inhabit."\cite{hhstatement}

Hence, rather than presenting all of the known physical characteristics or discoveries associated with a cosmic target, they emphasize using a visually compelling picture to inspire the viewer. The understanding is that the target audience will search relevant media, books, etc.\/ for more information.

The aim to captivate and inspire an audience requires an understanding of image-making: how images are read and digested by that audience as well as how images can be made to be compelling. These aspects fall under the purview of visual literacy, which uses  techniques such as composition and colour theory to form its visual grammar.  However,  research astronomers who are  image-makers also want to emphasize the scientific meaning in their images.   Sometimes I describe the endeavour to balance the visual impact with the physical meaning as a struggle between the culture of science and the culture of art.  

Noting that most of the readers of this article will be scientists, this paper focuses on introducing the reader to the role of the culture of art in astronomy public outreach images. A partial history of outreach images, \S\ref{history},  illuminates the constraints on image-making practices, their context, and outlines who participates in their production. A brief literature review, \S\ref{litreview}, highlights reflections by both astronomers and cultural theorists.  The motivation for using techniques from art is based on human physiology, \S\ref{perception}. The technical workflow for the construction of the images is outlined,  and image depiction categories defined, in \S\ref{construction}. The relevant visual art components such as composition and colour harmony, are introduced via the motivation for their use in  \S\ref{visart},. Examples of the retention of scientific meaning in the resultant image are presented in \S\ref{scimean}, and how  these images  feed back into scientific papers and experiments, including 3D virtual reality,  in \S\ref{feedback}. Future manifestations of visual outreach materials, \S\ref{future}, may target specific audiences and take on new forms, particularly in the art realm.  I conclude, \S\ref{conclusions}, that, given a struggle between cultures, both the culture of science and 
that of art win in astronomy public outreach images.   

This is not a comprehensive review of visualization techniques nor of our understandings of aesthetics, which have been covered in more detail by Rector et al.~\cite{travis2007} and Lindberg Christensen\cite{lars2014} respectively.  Rather, it is the perspective of a research astronomer who trained as a visual artist and who promotes visual literacy to both research scientists and the public.This personal point of view is illustrated mainly using images, produced with collaborators, from my own portfolio.

\section{A Sampling of Astronomy Image-making Endeavours: Setting the method.}\label{history}

\subsection{Film Photographers at Observatories}
\label{film}
Historically producing astronomy images for textbooks, magazines and newspapers was of some priority for observatories. The outreach images\cite{aaowebsite} and books\cite{malinBooks}  
by photographic scientist/astronomer David F. Malin for the Anglo-Australian Observatory provide examples of how successful this  endeavour can be. 

\begin{figure}[pb]
\centerline{\psfig{file=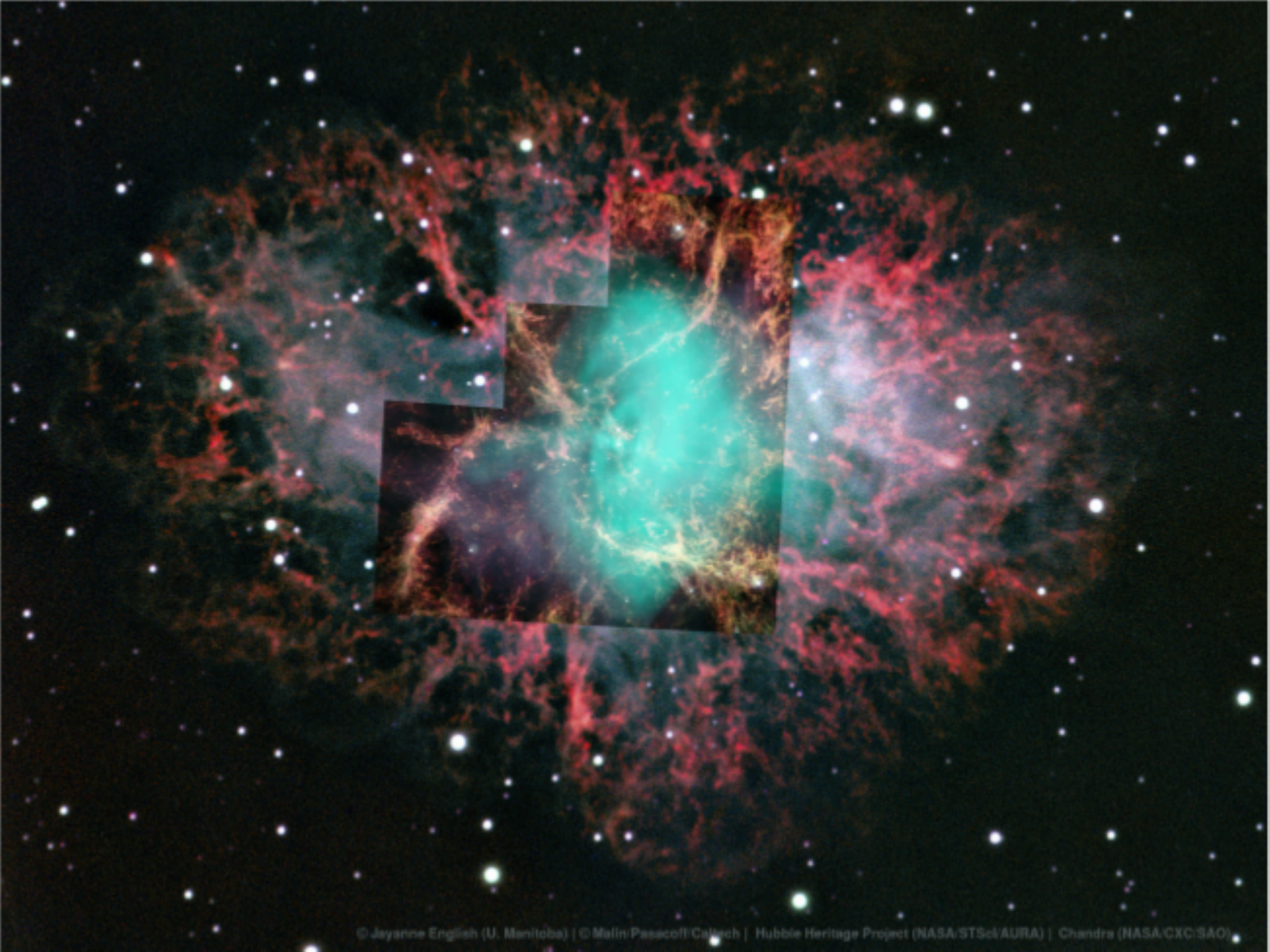, width=4.0in}} 
\vspace*{8pt}
\caption{Crab Nebula. The full field of view 
used Palomar 5-metre telescope photographic plates that match the range of wavelengths detected by human vision. Magenta in the outer regions shows the predominance of ionized hydrogen emitting at the red wavelength 656.3 nm (H$_\alpha$) diluted with emission at the blue wavelength 486.1 nm (H$_\beta$).  The chevron shaped inset was produced by the Hubble Heritage Team using the Wide Field/Planetary Camera 2 detector on HST. Only small segments of the human visual wavelength range  were sampled since the data were acquired  in order to characterize the elements and temperature of the gas. These data are assigned colours that support the scientific results (See Fig.~\ref{figCrabBlairHST}).  The central object is a pulsar wind nebula observed outside the visual wavelength range using the Chandra X-ray Observatory. It has been assigned turquoise in order to use a blue within the human visual range to represent the high energies associated with X-ray wavelengths. Visual grammar (\S~\ref{perception}) also plays a role in colour selection.\label{figMalEngCrab}}
\end{figure}

The first observatory photographic specialist to make colour photographs on a professional telescope was William C.~Miller (Hale Observatories; the Andromeda galaxy in 1958\cite{millerM31}).  Malin was the first person to to make true-colour additive (RGB) images\cite{malinVistas,malinElab} (1978). 
The technique involves obtaining black and white negative photographic plates for 3 separate wavelength ranges (e.\/g.~R (red) from 610 to ~700 nm; G (green)  from 495 Ð- 625 nm;  and B (blue) from 395 Ð- 500 nm) that are combined using colour darkroom techniques into a colour image. This mode of acquiring  data spanning a number of specified segments of the electromagnetic spectrum continues to this day, although the colourization of each range is now done digitally.  

Another continuing practice is that the black and white data were primarily acquired for astronomical research purposes, occasionally with some Director's Discretionary observing time 
 if supplementary data were required to produce a colour image.  
Indeed measuring specific physical characteristics, such as stellar temperature, impacted the standardization of wavelength ranges for filters, i.\/e.~passbands (\S~\ref{motive}).  Photographic specialists'  endeavours resulted in significant contributions to astronomical research. (E.\/g.~Malin's 
 discoveries of subtle structures in galaxies.\cite{malinShells})

However,  Malin's focus is currently on astrophotography for the layperson, who often asks ``Is this what my eye would see if I was in outer space?"  Indeed Malin's images approach the experience of human vision while contemporary digital examples of outreach images do not.  Our eyes are not sensitive to faint light and hence colour is drained from our ocular view of astronomical phenomena.   Malin's careful selection of emulsions and filters (spanning the human wavelength range),  
his calibration of colour,  and his various darkroom techniques\cite{malintech} 
results in images that reproduce what the eye would see {\it if} the human eye were as sensitive to photons as is photographic film.  That is, his images capture the intrinsic colours of the object as they would be seen by someone with superhuman sight. For comparison with colour selections made for modern outreach images, whose original data  do not match the human colour range,   Fig.~\ref{figMalEngCrab} combines Malin and Jay Pasachoff's  remastering of 1956 plates of the Crab Nebula\footnote {The Crab Nebula is a prototypical example of a supernova remnant. The core of a star, which is initially several times more massive than our sun, implodes. The collapsing stellar envelop smacks into the compacted core and is ejected due to Newton's 3rd law. The core becomes a neutron star while the envelope becomes the supernova remnant.} from a ground-based telescope 
telescope 
with 2 contemporary NASA space-based outreach images.  

\subsection{Instituitons Appropriating Images from the Scientist}
Around the mid-1990s many research astronomers had access to colour monitors and printers since their expense had dropped dramatically. While they were familiar with computer graphics software, astronomers did not have other skills relevant to making striking images for the public.  Nevertheless their work was disseminated in the popular media by the public communication wings of their institutions.  These images either were figures published in professional journals, created for communication with colleagues,  or were colour explorations of data for 
qualitative insights\footnote{For example, a figure from my PhD thesis was used as a postcard and season's greeting card in 1993 by The Australia Telescope National Facility as well as Fig.~1 in English et al.~2003, Astronomical Journal, 125, 1134.}.  

\begin{figure}[]
\centerline{\psfig{file=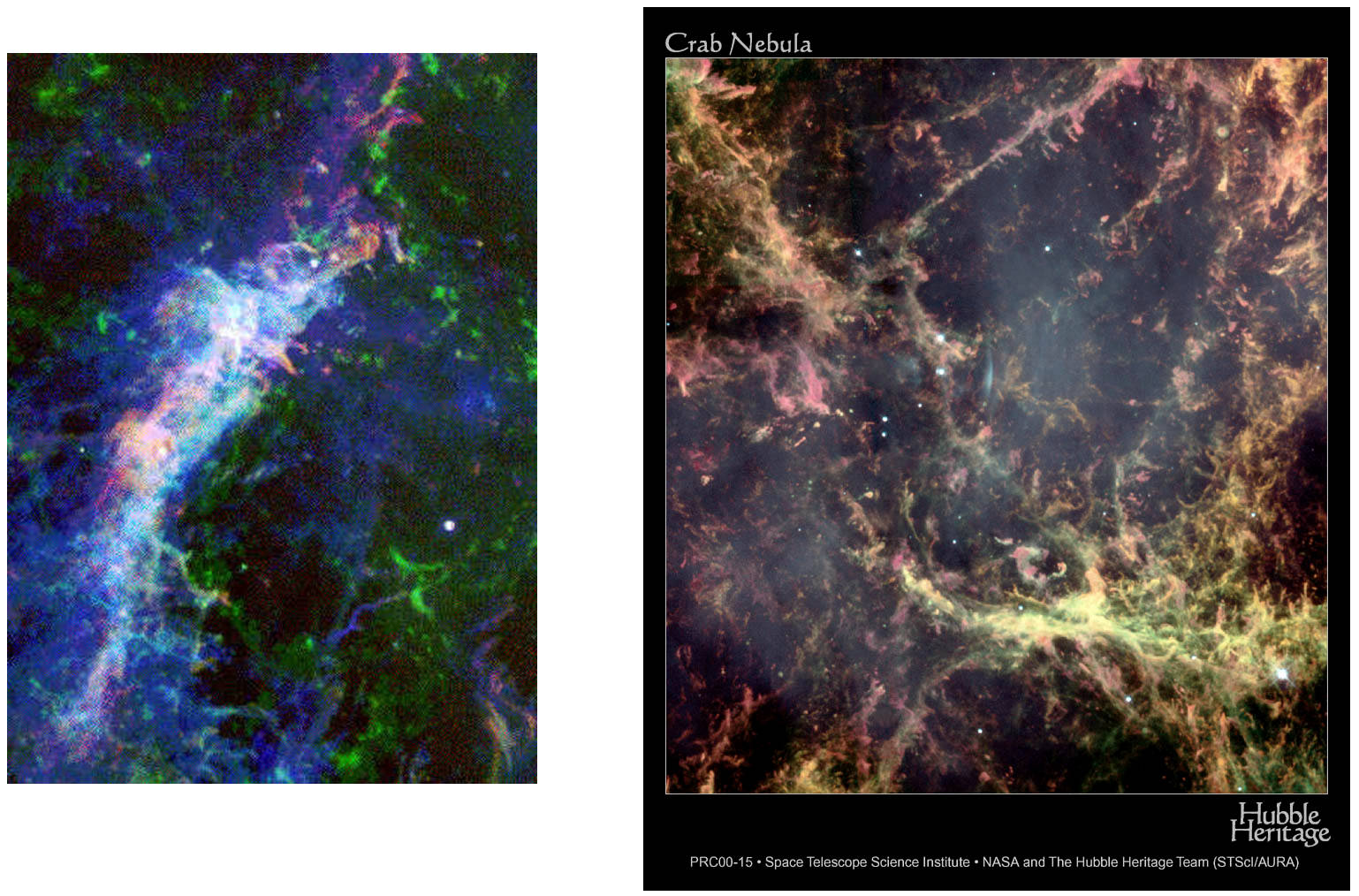, width=3.95in}} 
\caption{HST mages of the Crab Nebula. Although the fields of view are displaced from each other, the  black and white data were acquired for the same observing project and have identical characteristics. Left: A filament of the Crab Nebula.  Black and white HST data were assigned colours by William (Bill) Blair  to indicate the temperatures of the gas\cite{blair}
Right: The Heart of the Crab Nebula.
Techniques from visual art (\S~\ref{visart}) were employed by the Hubble Heritage Team to create spatial depth and a dynamic HST image. However,  these data were simultaneously assigned colours indicating the elemental composition of the gas. Green has been assigned to ionized oxygen, visually warm colours to ionized hydrogen, nitrogen and sulfur, and blue to the pulsar's synchrotron emission. (Details about the colour assignments 
are online\cite{crabfacts} while filters are described in \S\ref{motive}.). 
\label{figCrabBlairHST}}
\end{figure}

As with the historic professional astrophotography, the culture of science plays a significant role in assigning colours in these appropriated images.  Instead of looking like the Apollo 11 photographic view of the Earth, the astronomy images are more like a map of a classroom globe where Commonwealth countries are coloured pink. 
Scientists often find images with strong colour contrasts appealing, since their scientific discovery can appear visually significant, and therefore believe that these contour plot-like images appeal to the lay person as well. 

However,  this clarity requires a text legend. For example, the coloured HST digital images  of the Crab Nebula, such as  the left-hand side of Fig.~\ref{figCrabBlairHST}, disseminated on the web by William Blair, have captions such as: 

``See how the faint blue gas seems to envelop the the other colors? That tells us that the hotter, lower density gas surrounds the denser, clumpier gas seen in the other colors!"\cite{blair}

The physical information clearly takes precedence over the visual formalities in such images. Any set of colours that contrast strongly could be applied to the black and white data by the image-maker,  regardless of how they are read by the public.  Indeed most people read blue as a cool temperature  since it is used to represent ice or water in paintings, etc. 
Consequently the information in Fig.~\ref{figCrabBlairHST}  is not apprehended easily by a layperson, in spite of the physically reasonable color assignment.  Additionally, when transferred to  the public domain, the legends 
are, at best, buried at the end of an article and, at worst, stripped off altogether by editors or journalists. As a result the public viewer could see a meaningless splotch or, if the caption is too terse, learn something non-sensical.\footnote{I recall a newspaper clipping of the Cosmic Background Explorer  satellite map in which minute temperature fluctuations are coloured cyan for cooler and magenta for warmer measurements. The caption  read something like  ``astronomers discover the universe is blue and pink".} 

\subsection{Scientific Image-making Teams at Institutions: Merging the cultures of Art and Science.}\label{teams}
An additional mode of image-making that arose in the 1990s involved teams of scientists. Many were engaged at institutions, associated with ``big science" telescopes, 
which were  producing press releases aimed at the newspaper and broadcast media. Examples include institutes 
associated with NASA, the European Space Agency (ESA) and large ground-based observatories.  They treated the image-making team as a single entity, with the unfortunate result  that individual participants are usually treated as anonymous and are uncredited. Also, possibly because of the success of the images, official credit is often reduced to the largest agency in the list of collaborating institutions. 
Since these factors lead to the public believing these images are robotic snapshots (e.g.~``Hubble snaps a photo of...'' ), another intention of this article is to provide some information to help dispel that misunderstanding. 
 
The diverse compositions of image-making teams, which also produce text content and 
author their dissemination websites,  
could be seen at the 2003 ``Astronomical Outreach Imaging Workshop"\cite{baltoworkshop} in Baltimore.  The small-size end of the spectrum was represented by  Lars Lindberg Christensen and Martin Kornmesser from ESA/Hubble.  At the other end of the size spectrum was the several-person strong Hubble Heritage Team, working at Space Telescope Science Institute (STScI) and in direct communication with STScI's Office of Public Outreach (OPO). Original members Zolt Levay and Lisa Frattare continue to be the main image processors for both the OPO News office and the Heritage project\footnote{Initially 4 out of 8 members also worked in OPO, including  Levay and Frattare.  Levay has also been the Heritage Principal Investigator team lead since 2011. In 2016, the Hubble Heritage Team has 9 members comprised of image-makers, proposal planners, data analysts, outreach scientists, web maintainers, and interns.}.

If an institution employs an image specialist, a scientist works on the visuals for a press release with that  image-maker (e.g.~Robert Hurt (Spitzer Space Telescope); Kim Arcand (Chandra X-ray Observatory)), who either enhances an existing image or constructs an appropriate image from the astronomer's data. However,  these images are used to illustrate science deemed newsworthy - that is, the press release needs to present a new discovery, a significant step forward, a new mystery, or settle an area of controversy. 

The newsworthy criterion meant that in the 1990s many fascinating astronomical objects and aesthetically stunning astronomical imagery were not being presented to the public due to their apparent lack of news. 
Former STScI astronomer Keith Noll, inspired by Jeff Hester and Paul Scowen's powerful rendition of The Eagle Nebula\cite{EaglePillars} (1995) and the exciting public response it elicited, collected together some like-inspired colleagues at STScI to form the Hubble Heritage Project in 1998. Due to my familiarity with this team, I  use it as a primary example in this article.

The Hubble Heritage Project's directive 
was to disseminate striking images, regardless of newsworthiness,  that would pique the public's interest in astronomy. The team could apply for a small allotment of director's discretionary time on HST for interesting targets but only if  there was a strong scientific  case  for the observations.  Care was taken to choose targets and observing strategies that utilized Hubble's unique capabilities. However,  it was intended that the bulk of the data used for the releases would  come from the HST archive.  The team produced  content (e.\/g.~press release text, captions), authored their website, and coordinated with other NASA outlets (e.g.~ Hubblesite Newscenter; Astronomy Picture of the Day).  

Heritage team members collaborated on all these aspects,  yet  construction of the image  engaged the most team  members (\S~\ref{construction}).     Each target would have 1-2 lead image-makers, who would construct a series of renditions of the target, exploring colour options and compositions. 
A distinguishing feature of the Hubble Heritage images was that explicit aesthetic principles and scientific content both played a role in image production. Our goals included keeping the viewer engaged with the picture by creating, for example, spatial depth and colour harmony (defined in \S~\ref{colourwheel}). The team would discuss the options at length, exploring a number of iterations before settling on the final version.   As demonstrated by the right-hand image in  Fig.~\ref{figCrabBlairHST} the scientific meaning can be retained even when the image-makers are mindful of visual aspects\cite{crabfacts}.  Both images in Fig.~\ref{figCrabBlairHST} are based on data from the same observing project and contain similar filamentary structures. However,  this right-hand image is more photograph-like and hence the nebula seems more compelling as an object that the viewer can imagine existing in space.   

In order to have the support of our colleagues, both at STScI and the broader astronomy community, 
it was critical to interact with the astronomers who originally collected the target data.  We learned to strike a balance between visual creativity and scientific content that ensured that astronomers felt that their research was accurately represented and that their institute would be proud to be associated with the image.   
Since artificial textures, cut-and-paste compositions, and trendy colours were not appreciated,  we restricted our range of digital techniques (\S~\ref{manipulation}). 

The results were so satisfactory that  astronomers, whose data we were appropriating from the archives, requested permission to reproduce our public outreach images in their professional journal publications.
For example, the public outreach image of interacting galaxies NGC 2207 and IC 2163\cite{hh2207} was not only a photograph of the year in Life Magazine 
but was also used for the illustration of the special morphological features in interacting galaxies in the Astronomical Journal\cite{Elmegreen2000}.  Finally, communicating with the astronomers who collected the data allowed us to not only create images that resonated with the scientific results, it also  led to another mode of team image-making -- teams composed of scientific research colleagues.

\subsection{Scientific Teams of Research Colleagues: Beyond the single observatory}\label{colleagueteams}
Cleary Hester and Scowen are an early example of  an outreach image-making team composed of research astronomers working with data that they personally acquired for research.  Astronomer Travis Rector (U. Alaska) not only collects his data and converts these into images, he also  produces images for the international Gemini Telescopes consortium.   I produce images for the various research collaborations that I participate in scientifically.  Each of my projects has at least several 
 colleagues who work at internationally distributed institutions. Occasionally I may visit our principal investigator's institute for a week or 2 in order to develop the initial drafts of an image with the leader's input. Feedback is continually solicited from the research team and the materials for the overall outreach endeavour (e.~g.~a webpage, text for a press release) are also refined in collaboration. 
 
 There are several exciting aspects and advantages of creating images from data acquired for specific scientific questions.  
 For instance, we are not 
 restricted to data acquired by a single institute that is responsible for disseminating the resultant outreach image. 
 While the STScI 
 needs to promote HST data in its releases, 
 teams of research colleagues can explore many wavelength ranges, from X-ray through to radio energies.  This  frees the image-makers from expectations of colour  associated with strictly ocular vision. Additionally a large energy range  uses data from various telescopes, operated by different institutions, and our research data can be very current - not from an archive that is at least 1 year old.  Also we need not worry about whether the  entire community at a particular institute would be offended by our representation of the data collected at their telescope. (That said, although we can be more creative in our image-making approach, we still take care to represent the science accurately.)  Furthermore the pressure from a political or administrative body is essentially non-existent so there is less pressure to produce an image by an administratively set deadline.  Finally we can produce either a newsworthy image or simply a striking one, whichever is appropriate.  Regardless of type, the images produced on these projects can be disseminated through the outreach offices of one institute (e.g. STScI or the National Radio Astronomical Observatory or NASA, etc.) or, via a coordinated press release, through all appropriate institutes including the universities of all the scientific authors. 

\begin{figure}[]
\centerline{\psfig{file=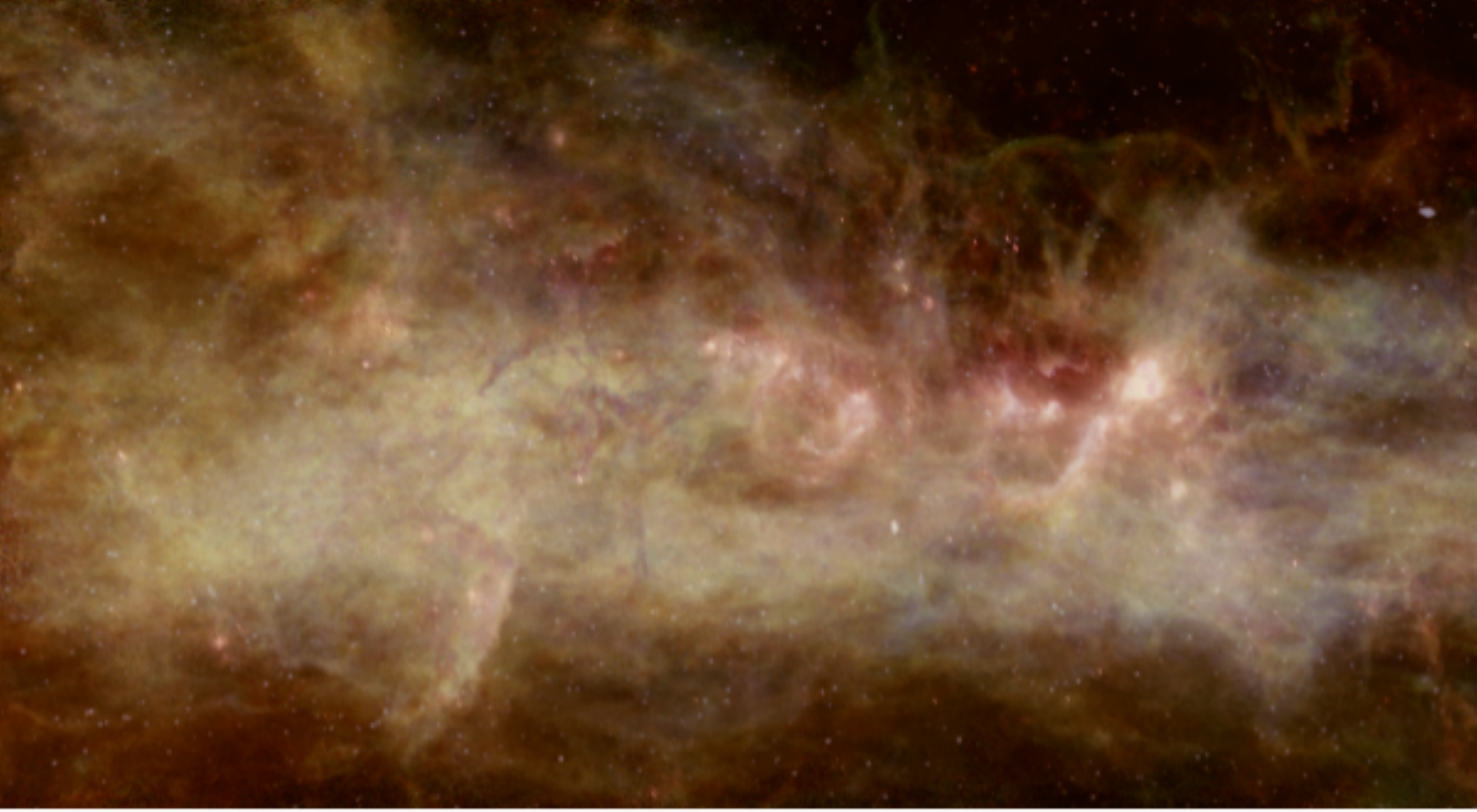, width=5.0in}} 
\caption{The Canadian Galactic Plane Survey view of the mid-plane of our Milky Way Galaxy. Emission from point-like  active cores of background galaxies was acquired with the Westerbork Radio Synthesis Telescope ($\lambda$ = 92 cm passband; assigned blue-grey). The Infrared Astronomical Satellite (IRAS) provides emission from warm dust ($\lambda$ = 60 micron filter; assigned pink).  Various colours were assigned to individual  velocity channels in DRAO 21 cm spectral line observations of HI. Since gas that is blue-shifted towards the observer was assigned a bluer colour, 
which dominates on the right hand side of the image, this image gives a sense of the rotation of the Galaxy. 
\label{figCGPSplane}}
\end{figure}

All of the factors above allowed me to combine data from 2-3 different telescopes to produce the first {\it photograph-like} radio maps.  These were done while I did research with Canadian Galactic Plane Survey (CGPS) consortium. While the image of the Cygnus Region,  in \S~\ref{scimean}, may be the most popular among my colleagues at the Dominion Radio Astrophysical Observatory (DRAO), the rather abstract appearance of the interstellar neutral hydrogen (HI) gas in the Milky Way's disk,  Fig.~\ref{figCGPSplane}, was genuinely satisfying to produce.  For an astronomical object to be easily approached by the general public, imaging teams at institutions have a tendency to select targets that have at least a recognizable, if not name-able, shape. Approximately spherical supernova remnants or planetary nebulae\footnote{In roughly 5 billion years our sun will eject its outer envelope in stages and the resultant shells will be ionized by the ultraviolet radiation from the core of our star.  The  core of our sun, and of stars initially with the approximate mass of our sun, will become a white dwarf star and this ionized envelope will be identified as a planetary nebula  (which has nothing to do with planets).} are popular.   Contrary to this approach,  
I mapped the velocity structure of the gas in our Galaxy without concern for whether the subject's morphology was familiar or not.

A spectral-line dataset acquired with a radio telescope (or with an optical/IR Integral Field Unit) is particularly rich since it is a 3D cube consisting of a 2D image of emission at each wavelength (i.e. frequency) channel for a number of channels.  Each channel consists of a specific range of wavelengths.
However,  in the case of classical optical observations a single image dataset spans a relatively broad range of wavelengths and different ranges are collected using 1 filter at a time. To acquire spectral information  non-image, 1D, multiple frequency data are collected using spectrographs.  In contrast, radio image data from multiple different channels are acquired simultaneously. Resulting in data cubes consisting of 2 spatial axes on the sky and 1 frequency axis, these are examples of 2D spectroscopy.  
One can tune the frequency range of radio observations to the HI spin-flip transition line at the rest wavelength of 21 cm. Depending on the target's motion, this single emission-line may be Doppler shifted throughout the overall frequency range of the data cube.  Thus one can trace the flow of the gas through its line-of-sight velocity ($ v $)  using the observed wavelength ($\lambda$) (or frequency ($\nu$))  and the speed of light in a vacuum: 
\begin{equation}
v = c \ \frac{\lambda - \lambda_{rest}}{\lambda_{rest}} = c\ \frac{\nu - \nu_{rest}}{\nu}. 
\label{dopplerv}
\end{equation}
 
During astronomical analysis a data cube is often collapsed along the frequency axis to form a 0th moment map of intensities or a 1st moment map for a velocity field\cite{CASAmoment2015} (see \S~\ref{perception} \& \ref{depth}).  Instead, for public outreach images, we harken back to the techniques initiated in the 1950s-1970s (\S~\ref{film}). Each channel 
is assigned a colour and combined with other channels,  as described  in \S~\ref{construction}.  {\it  Data from any wavelength regime can produce colour images using the same principles}, since image construction does not rely on how segments of the electromagnetic spectrum are obtained. 
However,  the scientific meaning embedded in an image will depend on the wavelength regime.  While optical filters may be used to acquire temperature differences, in Fig.~\ref{figCGPSplane} the colour assignments to the velocity channels gives a sense of the rotational motion of the disk of our home galaxy.

 \subsection{Citizen Scientists: Contributions from Non-professionals} 
 
 For completeness, we mention a few more imaging endeavours by research astronomers.  One mode consists of an astronomer interacting with a commercial entity to produce an outreach image, such as    Jean-Claude Cuillandre who interacts with Edizioni Scientifiche Coelum to create the Canada France Hawaii Telescope's (CFHT) ``Image of the Month"\cite{CFHT}. In another mode professional astronomers work with citizen scientists to produce images for research and professional publications.  R. Jay GaBany, has used his Blackbird Observatory to  support the research of David Martinez-Delgado (MPIA, IAC) et al.\cite{citsci1}. Finally, an archive, like the Hubble Legacy Archive, will occasionally hold an image-making competition, like ``Hidden Treasures", which invites the public to convert black and white images into colour.
  \ \\
  \ \\
\noindent In summary, this historical review of astronomy image-makers shows that the approaches developed in the era of black and white film astrophotography  are still used today to apply colour to data acquired in any energy range of the electromagnetic spectrum.  While contemporary images are composed with aesthetics in mind,  it is important to  the image-making teams, which are composed of astronomers, to represent the observations in a way that is respected by colleagues. This leads to the publication of public outreach images  in professional research journals as illustrations of scientific results.

\section{Literature: Reflections by astronomers and cultural theorists}\label{litreview} 
Not only are  contemporary, professional astronomers  reflecting on their process of making outreach images (\S~\ref{reflectAstro}), cultural theorists (e.\/g.~art historians) are also examining our practices(\S~\ref{culttheory}).\footnote{For historic overviews see references in \S~\ref{culture} as well as material by Malin\cite{malinBooks} 
and references therein. For general photographic techniques M.~Peres (2007)\cite{Peres2007}.}  The `literature' on the topic is not restricted to journal articles.  For example, the praxis of producing colour astronomy images is described by astronomers at conferences and  in professional newsletters, popular magazines, videos, and using web-based apps. The following examples of these dissemination modes include revealing perspectives from outside astronomy. 

\subsection{Reflections  by astronomers}\label{reflectAstro} 
A few articles on 
making contemporary 
colour images appear in professional research journals. For example, 
writing in Publications of the Astronomical Society of the Pacific, Richard Wainscoat and John Kormendy\cite{wainkorm1997} point out that the typical method of collecting scientific data does not permit making colour images that are true to human perception.  The paper for research astronomers by Travis Rector et al.\/(2007)\cite{travis2007},  in The Astronomical Journal,   details some of  the techniques outlined in \S~\ref{construction}.   Image-makers also publish in non-astronomy, academic journals. In RedFame's Studies in Media and Communication Joseph DePasquale et al.~(2015)\cite{pasquale2015} describe creating effective public imagery for X-ray astronomy while Kimberley Arcand et al.~(2016)\cite{arcand2013} describe selecting colours for  outreach images.

Other dissemination avenues that target research astronomers include the peer-reviewed Communicating Astronomy to the Public (CAP) Journal in which Lindberg Christensen, Hainaut and Pierce-Price published ``What Determines the Aesthetic Appeal of Astronomical Images?'' (2014)\cite{larsaesthetic}.  While the 6 parameters outlined by the authors  coincide, in general, with the image processing steps in in this paper's \S~\ref{construction}, \S~\ref{visart}
shows that their recommended prescription is only one possibility since  art techniques 
 lead to many alternative, and visually successful,  solutions.   

 Visual practices are also disseminated at conferences such as the CAP conferences presented every few years by the International Astronomical Union and in the education component of annual national astronomical society meetings.  I have reached a 
broader 
audience from science, industry and government 
at the Gordon Research Conference on Visualization in Science and Education.

Astronomers also write books for the public in which their methods, selections and motivations are outlined. Examples include those by David Malin\cite{malinBooks} (\S~\ref{film}), by Charles Robert O'Dell\cite{odell2003}, 
and by Travis Rector, Kimberly Arcand and Megan Watzke\cite{rectorColor}.  The latter, mainly illustrated by Rector's striking outreach images, outlines how telescopes collect data,  details how images are constructed, and explores the issue of aesthetics, touching on the topics in \S~\ref{visart}. 

The books and articles  on image-making in the popular press by journalists are too numerous to compile here. Also coffee table books (e.\/g.~G.~Sparrow's ``Hubble Legacy Edition''\cite{sparrow}, D.~H.~Devorkin  and  R.~W.~Smith's  ``The Hubble Cosmos: 25 Years of New Vistas in Space''\cite{devorkinSmith}) proliferate since NASA makes no claim to copyright, places images in the public domain, and does not collect royalties for the use of images.  Image-makers are interviewed for information in some media articles. Examples include a commemoration of HST discoveries\cite{astromag2015} in Astronomy Magazine (April 2015) and an explicit interview with Hester and Scowen on their Eagle Nebula HST image for Physics World\cite{hestscoWP}. Occasionally  astronomers themselves write for popular magazines -- an example is the piece in Sky and Telescope on the radio and infrared  image of the Cygnus Region (\S~\ref{sciexamples}) by myself and Russ Taylor\cite{EngTayCygnus}.    

Some magazines and TV programs with strong web components create very entertaining, as well as clarifying,  web-based  interactive applications.  The iconic HST Eagle Nebula is dissected at the PBS Nova  website\cite{EaglePBS} and  the viewer can reconstruct the Cygnus Region, to music, at Horizon Zero\cite{HorizonZero}.  

There are also online videos about image-making. Zolt Levay reflects on astronomy images during his TED-X talk about HST\cite{zoltTEDVid} and provides technical details in a video for SPIE\cite{zoltSPIE}, the international society for optics and photonics. Also, for the International Year of Astronomy 2009, astronomer Theresa Wiegert and I filmed a lecture and tutorials that explain the process in detail. The playlist, describing  the removal of noise and artifacts from black and white data through to applying cosmetic corrections to the final outreach image, is on YouTube\cite{cosmosYouTube}.  This  paper is based on more current renditions of the 2009 lecture, one of which was videotaped in 2016 and is available at the Voices from Oxford website\cite{cosmosOxford}.   

\subsection{Perspective of cultural theorists}\label{culttheory}
 Many viewers ask if astronomy outreach pictures are what one would see with their eyes if they were in space close to the object. The images  are not for technical 
 reasons (\S~\ref{manipulation}).  Nevertheless,  many cultural theorists focus on their compelling  realism.  An argument by Anya Ventura\cite{Ventura}\footnote{Ventura incorrectly 
 ascribes  ``false colour''  to the images, which is when a range of hues is applied to one characteristic, such as temperature, within one dataset. \label{false}} about  the apparent duplicity of astronomy images  is balanced by the perspective of Ryan Wyatt\cite{Wyatt} 
 that there is no objective image if the goal is to communicate.  

Leonardo, the 
respected journal of the International Society for the Arts, Sciences and Technology\footnote{A board member, Roger Malina (associate director of Arts and Technology UT Dallas), not only encourages interdisciplinary endeavours, he is an astrophysicist.},  has published a version\cite{larsaestheticleonardo} of Lindberg Christensen et al. (2014)\cite{larsaesthetic}. Two articles by art historians are of particular relevance to outreach image construction. Lee McKinnon's `Toward an Algorithmic Realism: The Evolving Nature of Astronomical Knowledge in Representations of the Non-Visible'\cite{leonardoMackinnon}, notes that humans are only one of the many 
agents acting upon outreach images and that 
visual conventions are moving toward a realism influenced more by 
image-manipulation software than film photography. Shana Cooperstein, in  `Imagery and Astronomy: Visual Antecedents Informing Non-Reproductive Depictions of the Orion Nebula' \cite{leonardoCooperstein},  attempts to explain the insistence on producing images that appear to represent visible phenomena from data that are not visible to the human eye. 

These themes also appear in the thesis\cite{Allain} of anthropologist Rosalie Allain who was embedded in the Astrophysics Group at Imperial College London for a couple of months. In ``Spectrum of Visibility: An Exploration of Astronomical Techniques of Visualization" she proposes giving equal weight to both human and non-human visuality, as well as to the visible and invisible.


Assessment of astronomy images by art historians appear in books like `Seen and Unseen. Art, Science, and Intuition from Leonardo to the Hubble Telescope''\cite{Kemp} by Martin J. Kemp, known for his column 
in the journal Nature.  He states that the translation of HST's mechanical `perceptions' into cosmic landscapes requires a level of contrivance  beyond that of a traditional landscape painter.  Elizabeth A.~Kessler, who interviewed the Hubble Heritage team, in ``Picturing the Cosmos: Hubble Space Telescope Images and the Astronomical Sublime''\cite{Kessler} proposes HST images resemble nineteenth-century landscape paintings and photographs of the American West.   
James Elkins 
instead believes the images depend on everything from science fiction book covers and movies to Maxfield Parrish works.  In his insightful ``Six Stories from the End of Representation: Images in Painting, Photography, Astronomy, Microscopy, Particle Physics, and Quantum Mechanics, 1980-2000"\cite{Elkins} he considers public outreach images to be eye candy that alienates serious art making from serious science.

Writer Evan Snider\cite{Snider} claims 
that the discourse on this topic is influenced by
a sociology article called ``Aesthetics and digital image processing: representational craft in contemporary astronomy''\cite{LynchEdgerton2015}, written in the 1980's by sociologist Michael Lynch  and art historian Sam Edgerton. They  interviewed research astronomer image-makers  and, intriguingly, watched them produce both scientific journal figures and {\it false colour mappings}\footnote{See Footnote~\ref{false} for description of false color.}  
for the popular media. They noted that astronomers reorient their images to what they perceive to be the aesthetic judgments of their audiences.   However,  the technique for producing public images (\S~\ref{intro} and  \S~\ref{manipulation})  has drastically changed from this 
simple colourization of contour plots.  Additionally contemporary image makers  find that their public outreach images can also be used  to illuminate the science in professional journals, demonstrating that their aesthetic choices 
currently do not change significantly for different audiences.



\section{Human perception and its relevance to astronomy images}\label{perception}
If the goal of the image-maker is to pique the interest of the general viewer so that they initiate their own journey of discovery, based around the subject that has been visualized (\S\ref{intropower}),   two components are required: scientific content and visual engagement.   It is the role of the image-maker to balance these.

However,  research-focused image-makers tend to act subconsciously with respect to the visual aspect of their work.   They are visual creatures in that their analysis tools privilege the use of the eye.
That is, astronomers routinely use visual, interactive computer software for measuring data;  they construct contour plots with extreme colour contrasts; and they manipulate  graphs to emphasize subtle deviations and to find trends.  Naturally they are affected by their everyday visual surroundings and exposure to art culture, which provides a range of influences on their visual products. 
Thus some visual literacy is unconsciously absorbed.

This article, on the other hand, endeavours to make visual construction and apprehension explicit. Knowing how one's audience {\it reads} an image can enable an image-maker to construct a more potent public outreach image or even scientific diagram. Additionally, since scientists are less familiar with these aspects of image-making, this paper aims to  
introduce these readers to visual literacy, rather than provide
a technical software manual for the production of outreach images. 

In analogy with reading and writing, there is a visual ``grammar". The arrangement, in this case, is of visual elements  and the method of arrangement incorporates the techniques of composition and 
colour harmony.  (Harmony is not an opinion but rather a definition, described in \S\ref{colourwheel}.)  Visual grammar  allows one to create an image that engages the viewer and retains their attention. Using visual grammar one can create spatial depth, emphasize intriguing fine-scale detail,  and deliver rich colours.  Often it can help one communicate some of the scientific content without total reliance on a legend.   In the culture of science legends are a sacrosanct means of associating information with the visualized data.  However,  editors of newspapers, magazines, blogs, etc. strip off legends, since they take up valuable space --- visual grammar applied directly to the image can help mitigate this loss. 

\begin{figure}[]
\centerline{\psfig{file=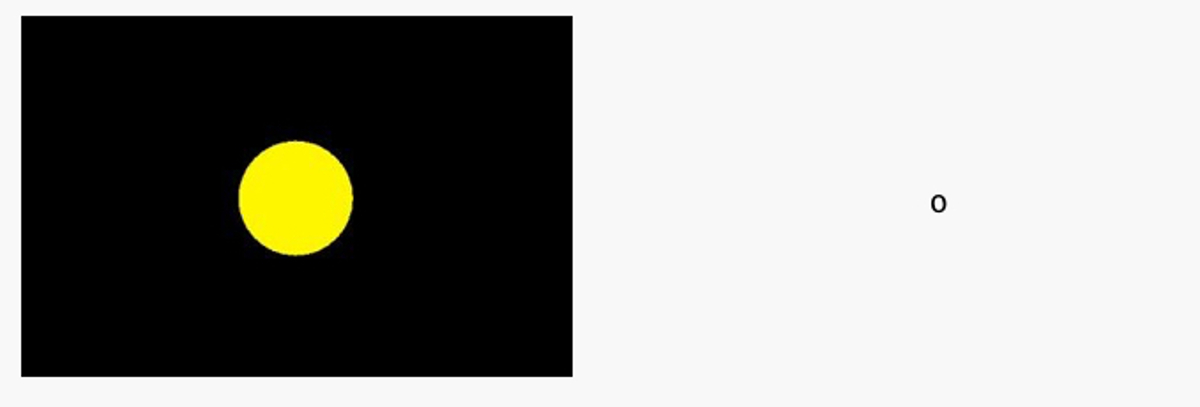, width=5.0in}} 
\caption{Afterimage of the colour yellow.  Stare 
attentively at the yellow circle on the left for 15 seconds and then look at the ring on the right.  What colour do you see surrounding this ring? (See text.)  
This illusion demonstrates that colours are not physical but an emergent property of characteristics associated with seeing light (e.g.~reflectance, luminance, wavelength) in combination with the eye-brain vision system.  Visual grammar can be used to take into account such responses to visual stimuli by human physiology, thus increasing  our engagement with an image (\S~\ref{visart}). \label{figafterimage}}
\end{figure}

 It is human physiology that not only plays a role in creating what we see but also determines our readings of imagery.  So let us start by demonstrating its strength.   Stare --- for 15 seconds --- at the yellow circle in  Fig.~\ref{figafterimage} and then look at the small black ring on the right while asking yourself what colour you see. When doing this experiment don't look around during the 15 seconds and don't 
 think about it theoretically.  Actually look at the yellow circle with attention.

What name would you give the colour you saw?   It is likely to be the purple-blue, associated with wavelengths around 460 nm (see right hand diagram of Fig.~\ref{figfilterCIE}).  Yellow and this purple-blue are complementary colours (described in \S~\ref{colourwheel}). The circle of purple-blue that you saw upon the black ring is an afterimage.  It does not exist physically --- the colour isn't painted with pigment around the ring.  Briefly, this illusion results because the receptors in your eye that respond to yellow light get ``tired" while the ones that respond to blue-purple continue to function.\cite{eye}  

What is relevant to this paper is that physiological processes like this one impact every stage of image production and the reading of the image.  For example: \\
--- If an image-maker has any colour other than achromatic white, black or neutral grey on the background of their monitor they will not be assigning to their data the colours that they think they are assigning. The surrounding context has a direct impact on a perceived colour, and a complementary colour can be spontaneously generated by a surrounding colour, as in Fig.~\ref{figafterimage}. 
E.\/g.~if a monitor background is green 
the eye plus mind will spontaneously generate and impose the complement to green (which is magenta) on any nearby colours. Thus yellow will appear to be orange-ish, since it combines  
with the magenta afterimage (\S~\ref{colourwheel}).    However,  this orange 
will not appear  to another viewer of the outreach image, unless that viewer has the same green background on their monitor. (See contrasts in \S~\ref{contrasts}.) \\
--- The image-maker can choose colours that resonate with physiological responses in order to strengthen the impact of their images and to highlight their salient scientific features.  Indeed the image-maker can pick their favourite colour and do the experiment in Fig.~\ref{figafterimage} in order to know what colours to aim for in order to create an harmonious image (defined in \S~\ref{colourwheel}.)\\
-- How an audience reads a particular colour in the final image is also affected by whether surrounding colours are light or dark, redder or bluer, fill a large area or not.   These  factors can determine whether a colour comes forward or recedes.  
Additionally, the placement of a colour within in the composition can determine how the viewer's eye moves around in the image (e.\/g.~\S~\ref{composition}). 

\begin{figure}[h]
\centerline{\psfig{file=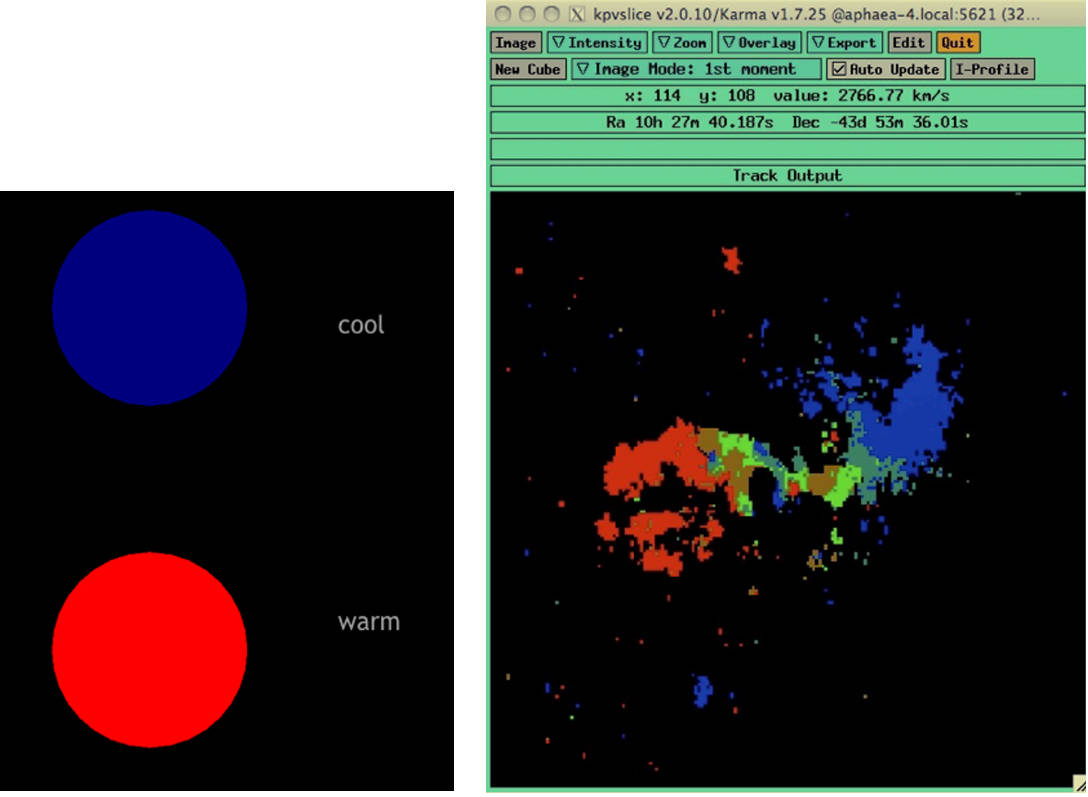, width=4.25in}} 
\caption{Warm and cool colours used in velocity maps.  Left: Which circle appears to jump forward?   They are the same diameter,  yet the red circle appears larger and hence in the foreground. In terms of visual grammar 
typically reds, oranges and yellows are called ``warm'' colours while 
blue and blue-greens that sink to the background are deemed ``cool'' (\S~\ref{depth}).  
Right:  A 1st moment 
velocity field map of NGC~3256, a rotating system of merging galaxies. This map uses a naive colour table in which red represents red-shifted emission from receding HI gas and blue for blue-shifted emission from approaching gas.  Visual perception does not support this representation since the red appears to be approaching. See \S~\ref{depth} for an alternative representation. (Australia Telescope Compact Array radio telescope observations rendered using the Karma Visualization suite.)
\label{figwarmcoolballs}}
\end{figure}

Physiological effects are generally ignored during analysis of data and in figures used in research publications, where legends are provided that associate a colour with a physical parameter. In the example in Fig.~\ref{figwarmcoolballs} the readers may ask themselves which of the identically sized balls appears closer or larger and answer that the red ball appears to jump {\it towards} the viewer.  
In contrast to this Doppler shifted (Eq.~\ref{dopplerv}) emission from gas that is {\it receding} from the viewer is by definition redshifted.  Thus the naive colour table used for velocity maps (e.g. right side of Fig.~\ref{figwarmcoolballs}) assigns to physically receding gas a red hue that is perceptually approaching the viewer. This contradictory rendition, at the very least, undermines the intended message if it is appropriated and presented to a public audience.  As well as elaborating on colour and composition,  \S~\ref{visart}  offers a way to mitigate this dilemma in communication.

\section{Constructing Public Outreach images: Technical Aspects}\label{construction}

Recall that compelling public outreach images can be constructed from any kind of telescope data and from any wavelength regime. Indeed, since different scientific observations show different physical processes 
an image-maker may wish to highlight a combination of phenomena in their images, as I did in Fig.~\ref{figCGPSplane}. 

One may also  combine data  that differ in their ability to show detail --  that is data with different spatial resolutions. Astronomical ``resolution'' characterizes the ability to distinguish between objects. For unresolved point-like targets their radiation is spread into  a 3-D distribution of intensity as a function of radius.  In the optical regime this forms a Gaussian `cone' parameterized as the ``point spread function''. More complex in the radio regime, this ``resolution element'' is the ``synthesized beam''. The resolution can be quantified using their 
Full-Width Half-Maximum (FWHM)  of such ``resolution elements''.  
It is not necessary to degrade high resolution data 
(whose resolution element has a smaller 
FWHM) to match low resolution (larger 
FWHM) data, since differences in resolution can be used to visually ``detect" objects (e.\/g.~the cores of active galaxies; see Cygnus Region  
(\S~\ref{sciexamples})). 

\begin{figure}[]
\centerline{\psfig{file=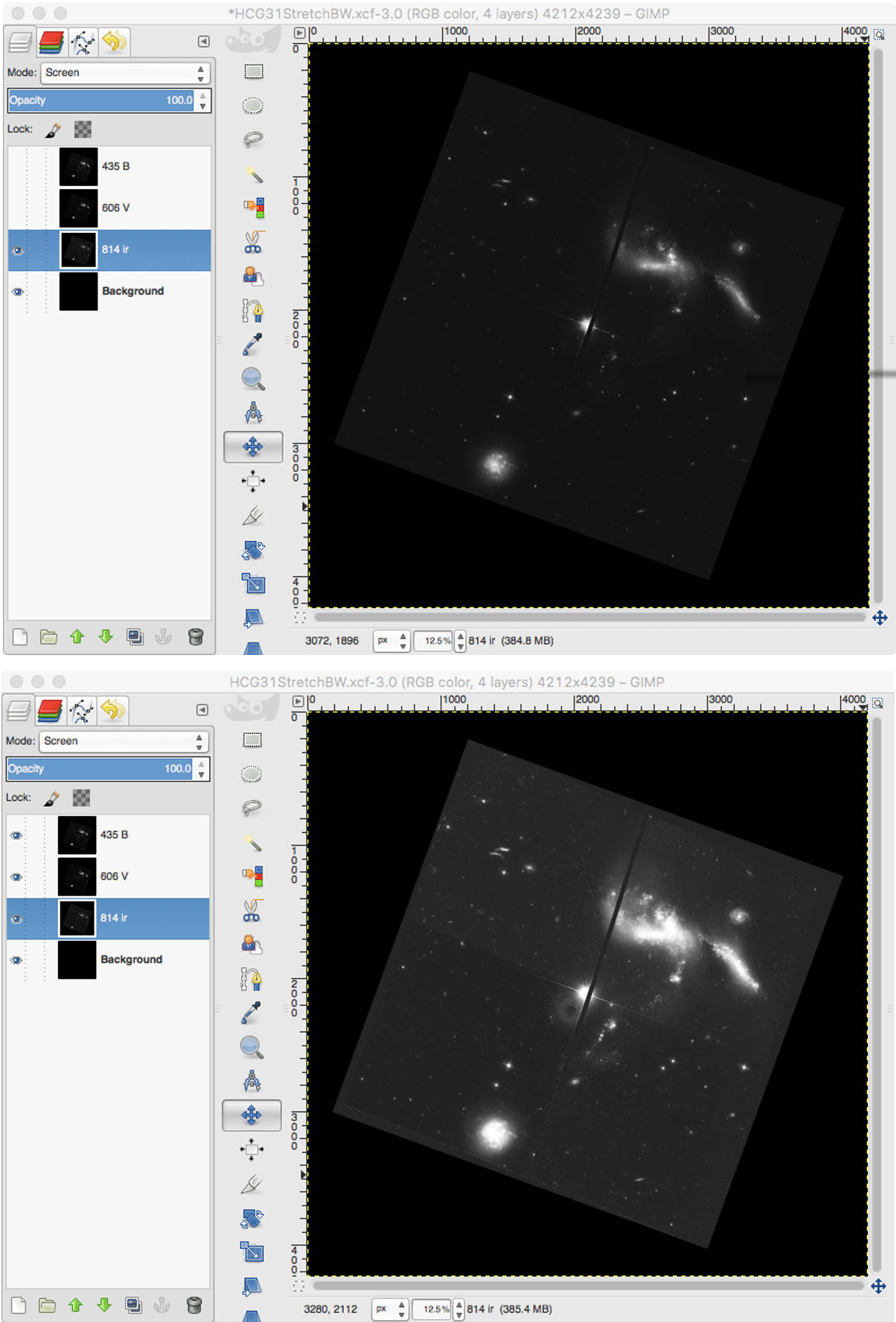, width=3.45in}}
\caption{The construction of the Hickson Compact Group 31 outreach image -- layering black and white images. The  layering schema is shown in the left-hand panel of each GIMP\cite{GIMP} window. Top:  Display of a stretched single HST filter in the Near-IR centred at 815 nm (F815W; \S~\ref{construction}: Step~\ref{stretches}). Bottom: Display of 3 HST filters (F439W, F606W, F815W), with adjustments (\S~\ref{construction}: step~\ref{layers}). The blending mode is shown on the left in the layers dialog box. Note the ``ghost'' ring near the centre, caused by scattering in the telescope optics by the bright foreground star..\label{fighcg31a}}
\end{figure}

\begin{figure}[]
\centerline{\psfig{file=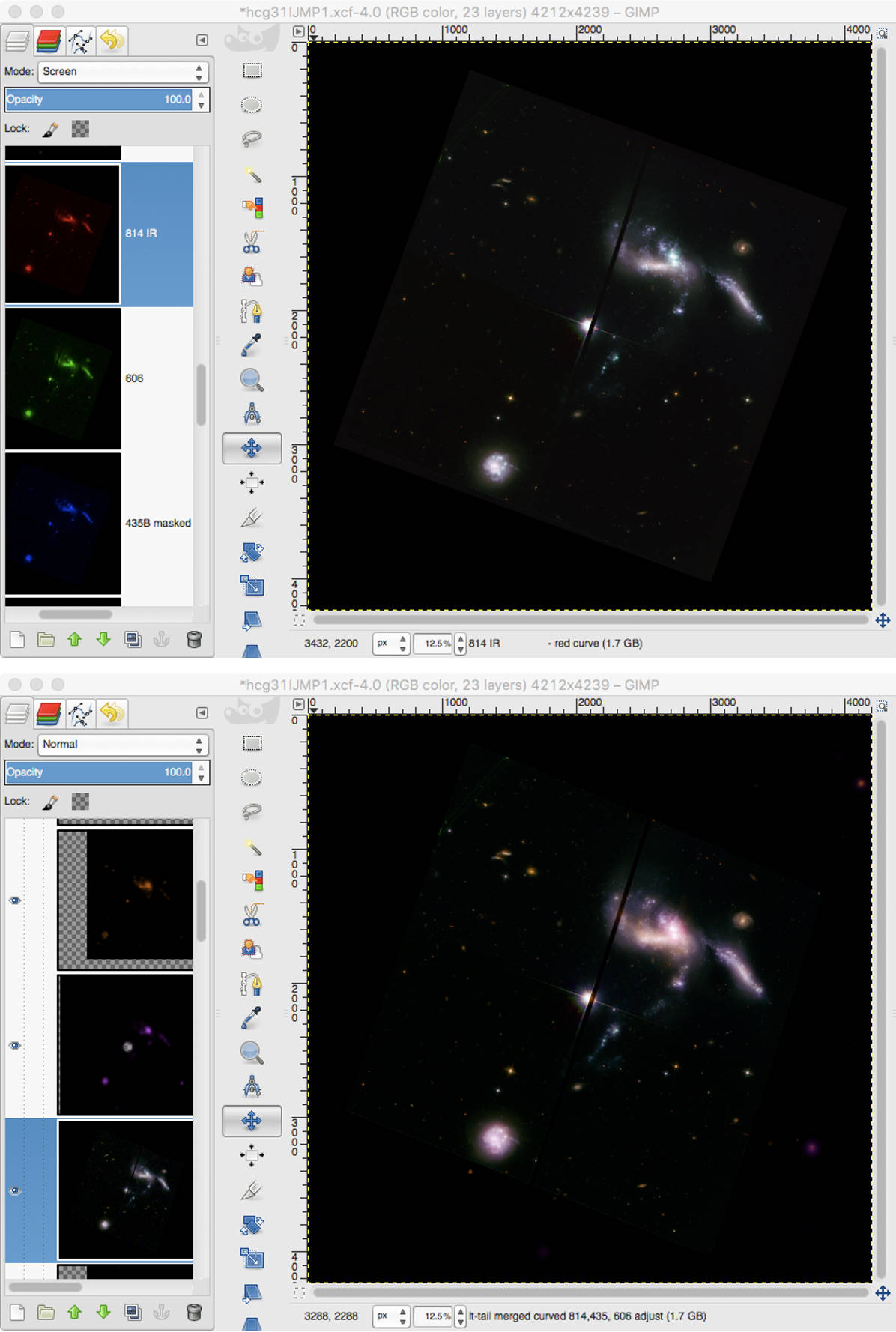, width=3.45in}}
\caption{The construction of the Hickson Compact Group 31 outreach image -- colour. Top: Assigned colours (\S~\ref{construction}: step~\ref{colourize}) are shown in the layers dialog box on the  left-hand panel of each GIMP\cite{GIMP} window.. Adjustments to intensities and masking of the ghost have been applied. Bottom: Mid-IR data from the SST (centred at 8 microns) and GALEX Near-UV data (227 nm) were also assigned colours and screened in with the combined HST data.  The foreground star's UV emission was
desaturated.  See Fig.~\ref{fighcg31final} for the final image.\label{fighcg31b}}
\end{figure}

\begin{figure}[]
\centerline{\psfig{file=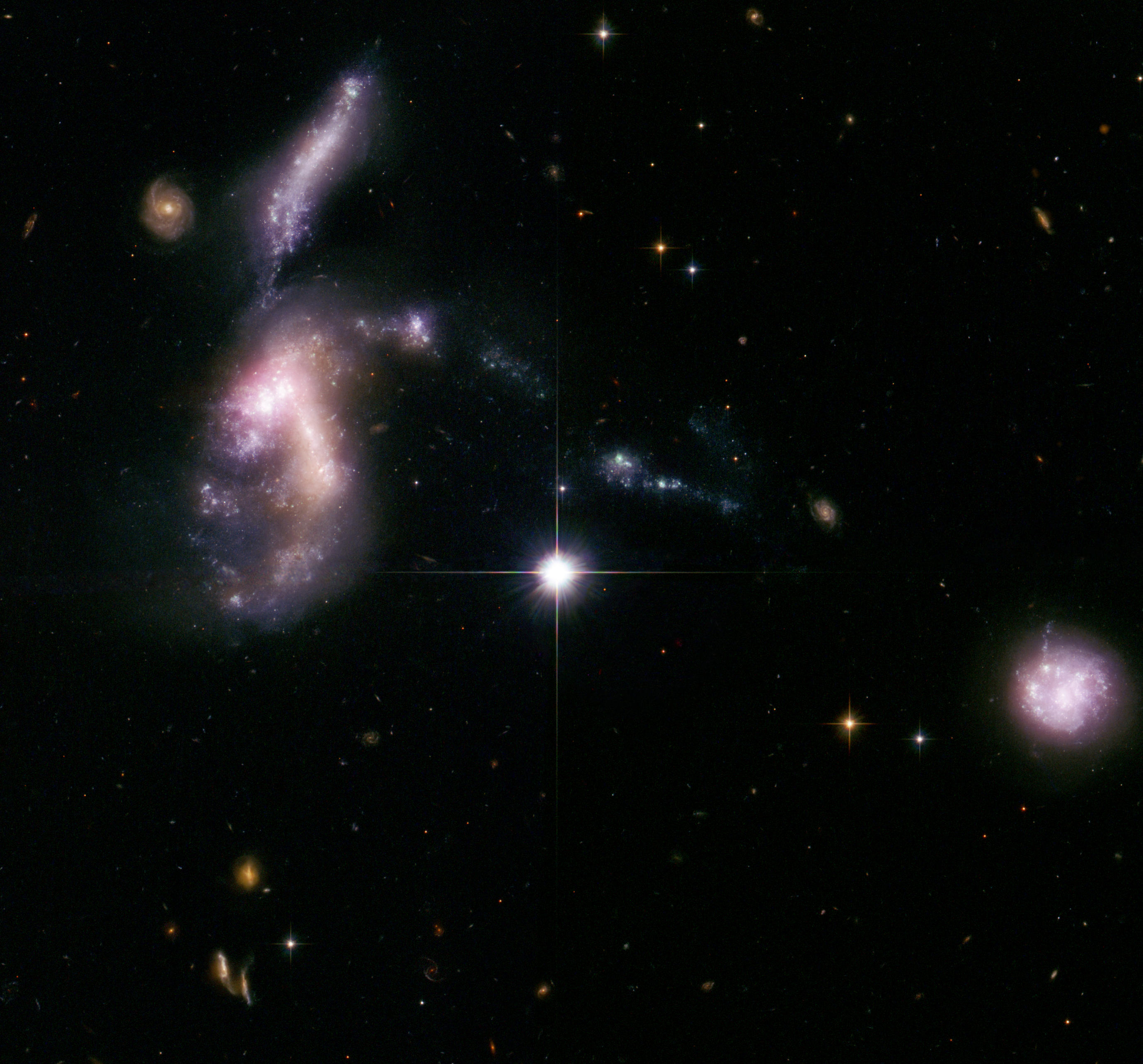, width=3.0in}}
\caption{HCG 31 outreach image\cite{releaseHCG31}. All layers in Fig.~\ref{fighcg31b}, bottom,  were combined (\S~\ref{construction}: step~\ref{combine}), the image composed (\S~\ref{construction}: step~\ref{compose}), and the foreground star was manipulated, to mitigate artifacts caused by removing the ghost (\S~\ref{construction}: step~\ref{cosmetics}).  Additionally the Hubble Heritage Team contributed to this image by  incorporating data from HST's Wide Field and Planetary Camera 2 detector to fill the gap between cameras in the Advanced Camera for Surveys detector.
\label{fighcg31final}}
\end{figure}

The following  ``workflow" steps for constructing an image are the same regardless of the wavelength regime of the data or the physical process that one may want to illuminate.  They are illustrated in Fig.~\ref{fighcg31a}, \ref{fighcg31b}   \& \ref{fighcg31final} using Hickson Compact Group 31\footnote{HCG 31, an association of dwarf galaxies,  fits inside the volume circumscribed by the diameter of our own Milky Way Galaxy.  They are close enough together to be interacting gravitationally with each other, resulting in intriguing tidal distortions of the stellar and gaseous structures.}  
Constructed by the author with feedback from co-investigators,  as in \S\ref{colleagueteams},  versions of Fig.~\ref{fighcg31final} have been published in a professional journal\cite{hcgGallagher}.  Additionally it was part of a NASA press release\cite{releaseHCG31}. 
Challenging to construct, it incorporates data from 3 space-based observatories: the Hubble Space Telescope (HST), the Spitzer Space Telescope (SST) and the Galaxy Evolution Explorer (GALEX).  The Hubble Heritage Team contributed at the last (cosmetic) stage. 

To combine and assign colours to 
individual images one uses an image manipulation package that employs a ``layering" scheme, such as Photoshop\cite{photoshop} or Gnu Image Manipulation Package (GIMP)\cite{GIMP}.  Layers of images are combined via algorithms called ``blending modes". The optimal mode is `screen',  which stacks images so they appear to be sheets of see-through transparencies. (If one decomposes a photograph into RGB channels and wishes to recombine them, `screen' will recover the original image while `addition' will not.) More detail is provided in Rector et al. 2007\cite{travis2007}. 

 \ \\
\noindent{\bf Workflow} (Figs.~\ref{fighcg31a}, \ref{fighcg31b} \& \ref{fighcg31final}): 
\begin{enumerate}
\item Process the individual data sets using 
professional astronomy software tools and techniques designed 
for scientific measurement purposes. Attempt to mitigate the  artifacts and noise. \label{process} 
\item Redistribute the intensity levels of each data set (called `stretching') using professional astronomy analysis display software. Save the displays of stretched data as individual greyscale images in common, uncompressed digital photography formats (e.\/g.~PPM, TIFF).\label{stretches}
\item Layer the greyscale images in an image manipulation software package.  Employ an  algorithm for combining layers so that 
each layer 
contributes to the displayed 
image. In each layer adjust intensities, attempt to further mitigate noise, and mask out saturated objects
and instrument artifacts (e.\/g.~optical `ghosts').\label{layers}
\item Assign colours to the the individual layers. Per layer adjust both intensity and 
hue to emphasize features relevant to the science content. \label{colourize}
\item Copy and export the displayed image into a single image in uncompressed format.  Adjust intensities and colours of this resultant image.\label{combine}
\item Adjust the composition via orienting and then cropping the image.\label{compose}
\item Mitigate noise and other artifacts using cosmetic adjustments, such as cloning and masking.\label{cosmetics}
\end{enumerate}


Unlike the production of a fine art photograph, this process involves many people as well as a number of complex instruments, devices and software programs, all referred to as ``actors" in cultural theory. The practices of the Hubble Heritage Team provide an example of how the above procedure can be applied
by a number of actors. Team members each take a different role on any given image, creating " an astronomy ``photograph" processing pipeline of sorts.  One team member handles the data in Step~\ref{process} ---preparing it, producing a mosaic of multiple pointings and fields of view, combining exposures from the same filters, etc. Another team member, in the role of image-maker (i.\/e.~image processor) performs Steps~\ref{stretches},   \ref{layers}, \ref{colourize}, \ref{combine}, and \ref{compose} ---   making a few color compositions to be discussed by the team.  Steps \ref{colourize} - \ref{compose} are reiterated  until the team settles on the composition. Finally  another member performs Step~\ref{cosmetics}, cosmetic cleaning and noise reduction. In this manner each person improves their skill at specific tasks.  

The workflow moves the data from an esoteric visual realm understood by physicists into a culture of images that is comprehended by a broad public. To illustrate this I'll use P. Galison's terminology for visual depictions, introduced in ``Image and Logic: A Material Culture of Microphysics"\cite{Galison1997},  and which have been adopted and extended by cultural theorists.  His category of western {\bf ``image tradition"} refers to pictorial representations that appear to ``preserve the form of things as they occur in the world".  At the other end of the depiction spectrum is the {\bf ``logic tradition"}\footnote{The word logic refers to logic circuits used in physics experiments.}, which refers to images created by devices that count, such as Geiger-Muller counters.   A characteristic of  the logic category is that quantitative information accumulates over time, and statistically, to produce a picture. 

Our imaging instruments on telescopes collect image data that lay towards 
 this logic tradition end of the spectrum.  For example, individual pixels count the number of photons over an exposure time in our charge-couple device (CCD) cameras, such as those on HST.  Statistically we increase the signal-to-noise ratio by aligning multiple exposures and combining them together.  The logic tradition category is particularly relevant to radio telescope arrays --- they synthesize a large mirror using the rotation of the Earth and only after a period of time can the morphology of the target 
be reconstructed. Furthermore, since reconstruction is done mathematically using a Fourier transform, more than 1 statistically valid image can be produced from the data. 
Starting at the Step~\ref{process} of our workflow we have a logic tradition type of image and the subsequent  steps attempt to make an outreach image that is in the western image tradition.  That is, the end goal is an image that is sufficiently naturalistic that it convinces the viewer of the existence of the depicted astronomical phenomena. 

An additional goal of this endeavour, and equally important, is the aim of retaining scientific content.  It is their investigative and discovery 
potential  that make these images relevant in both the scientific and public realms (\S~\ref{scimean}). 

\section{Techniques from the Visual Arts: Infusing ``image tradition'' representations with power.}\label{visart}

While a scientist finds enormous appeal in a scientifically clarifying contour plot of their data, a public viewer finds it hard to be engaged with such figures (e.\/g.~\S\ref{future}). Not only is the relevance of each contour level a challenge to digest, 
the figure provides nothing to distinguish whether the subject is biological, geological or astronomical. On the other hand public  engagement can be produced by 
employing techniques common in the culture of visual arts, particularly those that account for human perception (\S~\ref{perception}). These  techniques can provide the means for converting logic tradition maps into convincing western image tradition depictions of phenomena (\S~\ref{construction}). 

\subsection{Motivation for Using Art Visual Techniques: An Example}\label{motive}

\begin{figure}[]
\centerline{\psfig{file=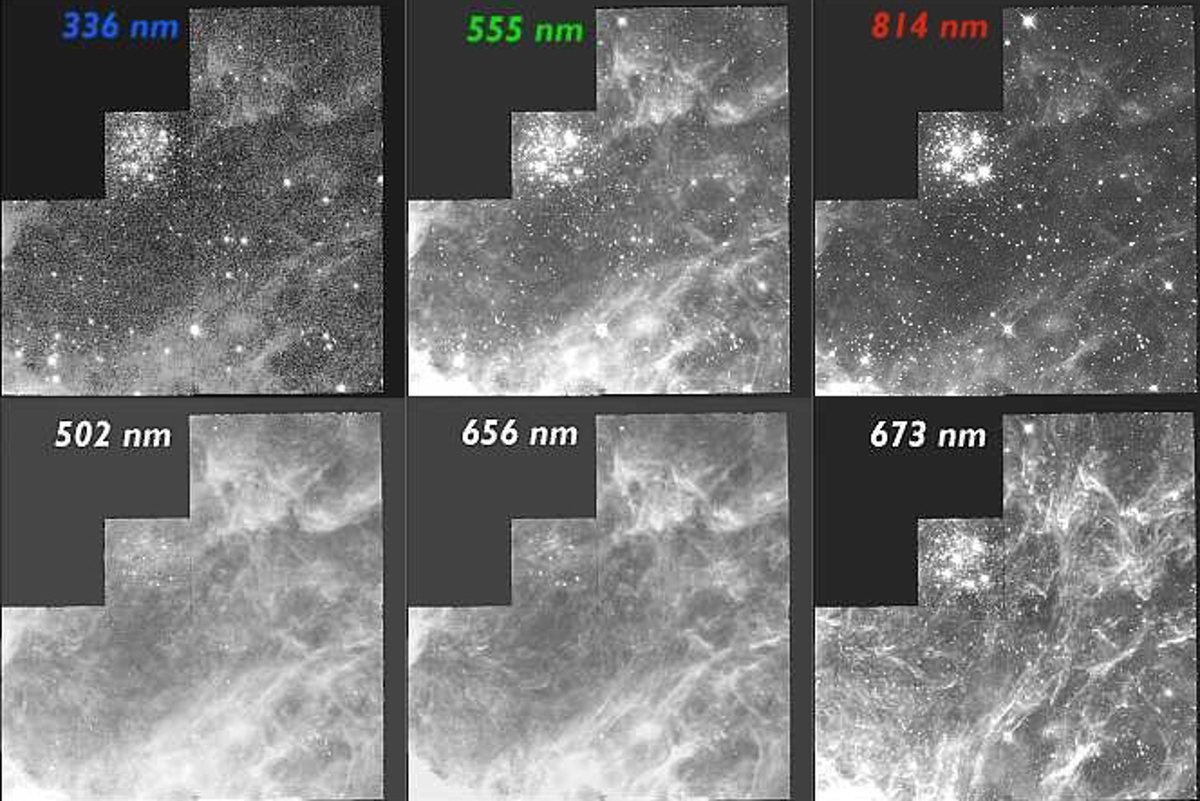}} 
\caption{Hodge 301 in the Tarantula Nebula though various filters.  Top row:   Broadband filters with effective central transmission wavelengths close to 336 nm (UV), 555 nm (optical) and 814 nm (near-IR) and filter FWHM widths of approximately 37 nm, 122 nm, and 176 nm respectively.  Bottom:  Narrowband filters with effective transmission wavelengths  in the optical regime close to 502 nm, 656 nm and 673 nm with filter FWHM widths of approximately 2.7 nm, 2.2 nm,  and 4.7 nm respectively. The left filter transmits  
emission from ionized oxygen [O III], the middle filter from H$_\alpha$, and the right from ionized  sulphur [S II].  Note that the last two filters trace very different structures although, referring to Fig.\ref{figfilterCIE},  to the human eye they would appear the same colour. \label{fighodge301}}
\end{figure}

\begin{figure}[]
\centerline{\psfig{file=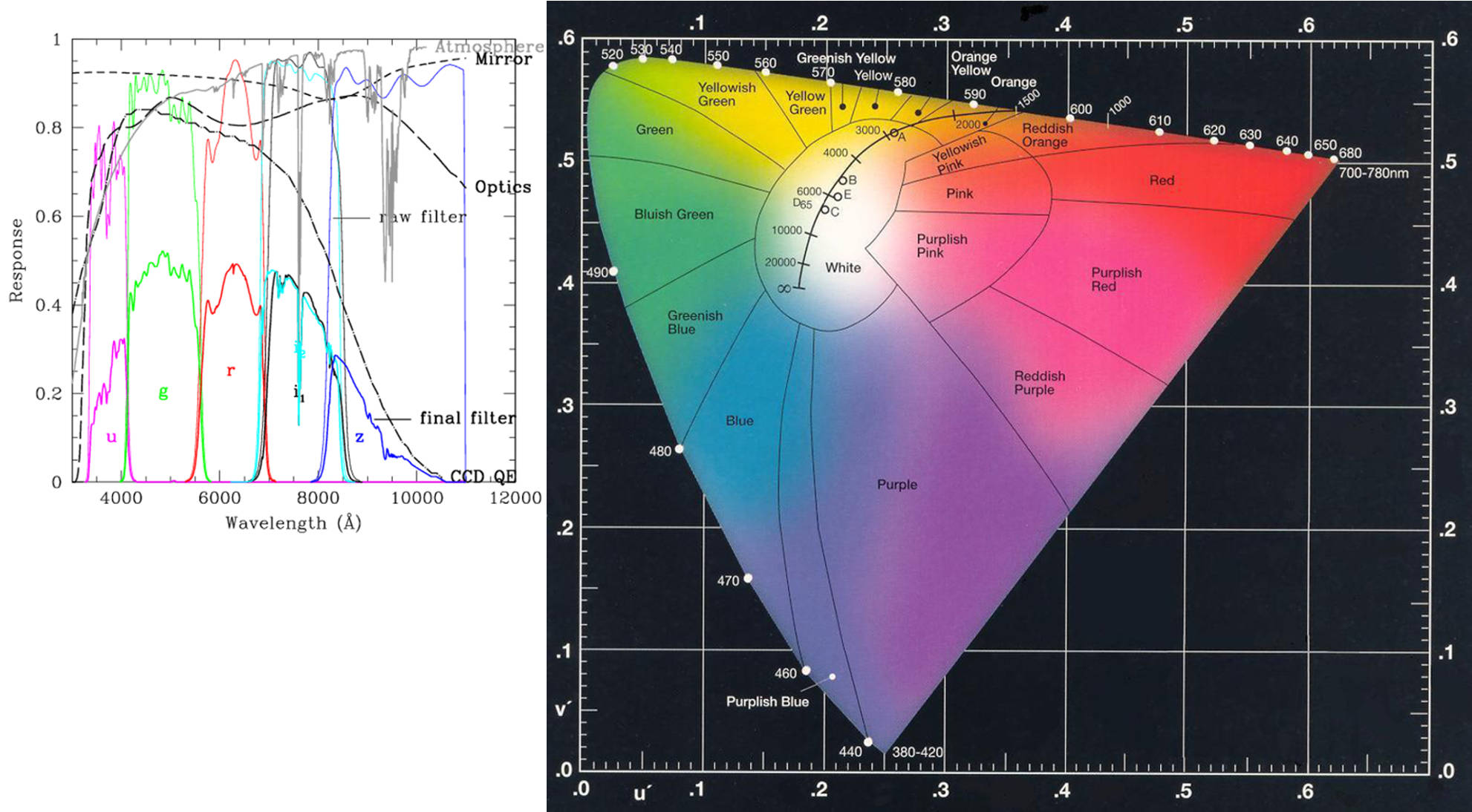,width=6.0in}} 
\caption{Astronomical Filter Transmission compared to Human Colour Perception. Left:  The response (1 = 100 \%) at CFHT's MegaPrime detector of  photons transmitted through a standard filter set. (See handbooks for CFHT\cite{cfhtfilters}, HST\cite{crabfacts}, etc.\/). Right:  A chromaticity diagram based on human colour perception produced by the Commission Internationale de l'Eclairage (CIE)\cite{cieinfo}. The x and y axes are related to hue and purity. The colours in the centre depict the gamut of normal human vision. 
The numeric  label encircling the locus around these colours specifies 
their wavelength in nanometers 
(1000 \AA\ = 100 nm).  Comparing these numbers with the values on the x-axis 
on the left shows that the majority of  perceived colours are transmitted through one  (g) filter. 
\label{figfilterCIE}}
\end{figure}

As stated in \S\ref{construction} the collected data are monochromatic and hence are black \& white when displayed, as in Fig.~\ref{fighodge301} of Hodge 301 
This example uses 
segments of energy in  the EM spectrum that are distributed between the UV and near-IR domains. Each different energy range is selected using filters.  
Fig.~\ref{figfilterCIE} shows transmission curves of a standard set of broadband filters.  The percent transmission of photons of certain a wavelength depends on  
the combination of the filter plus telescope system (CFHT)\cite{cfhtfilters}.  On the right  is the Commission Internationale de l'Eclairage (CIE)\cite{cieinfo} mathematically defined colour space used in studies of human colour perception.   Note that the photons that we perceive as violet, purple, blue, turquoise, green and yellow are passed through a single filter (`g', which is a standard filter at many telescopes) and thus the black \& white image from this filter contains most of the colours that we see!  We cannot distinguish between these colours using broadband filter images, which are the most common  type of CCD data images. 

As with Hodge 301, astronomy images  often include 
some data from outside the range of human vision.  For example a scientific visual narrative may use broadband data in the near-infrared to show the spatial location of older stars and narrowband optical data of H$_{\alpha}$ emission to pinpoint where gas is heated by newly formed stars\cite{combgalwebsite}.  Images like 
those in Fig.~\ref{fighodge301} contain such a mix of broad and narrow passband data\cite{hodgefilters}. This example  is complicated by the fact that the transmission peaks of 2 filters  reside in the same colour region (i.\/e.~red) in the CIE diagram. So the question becomes ``how does one map the selected passbands to perceived colours?''

\begin{figure}[]
\centerline{\psfig{file=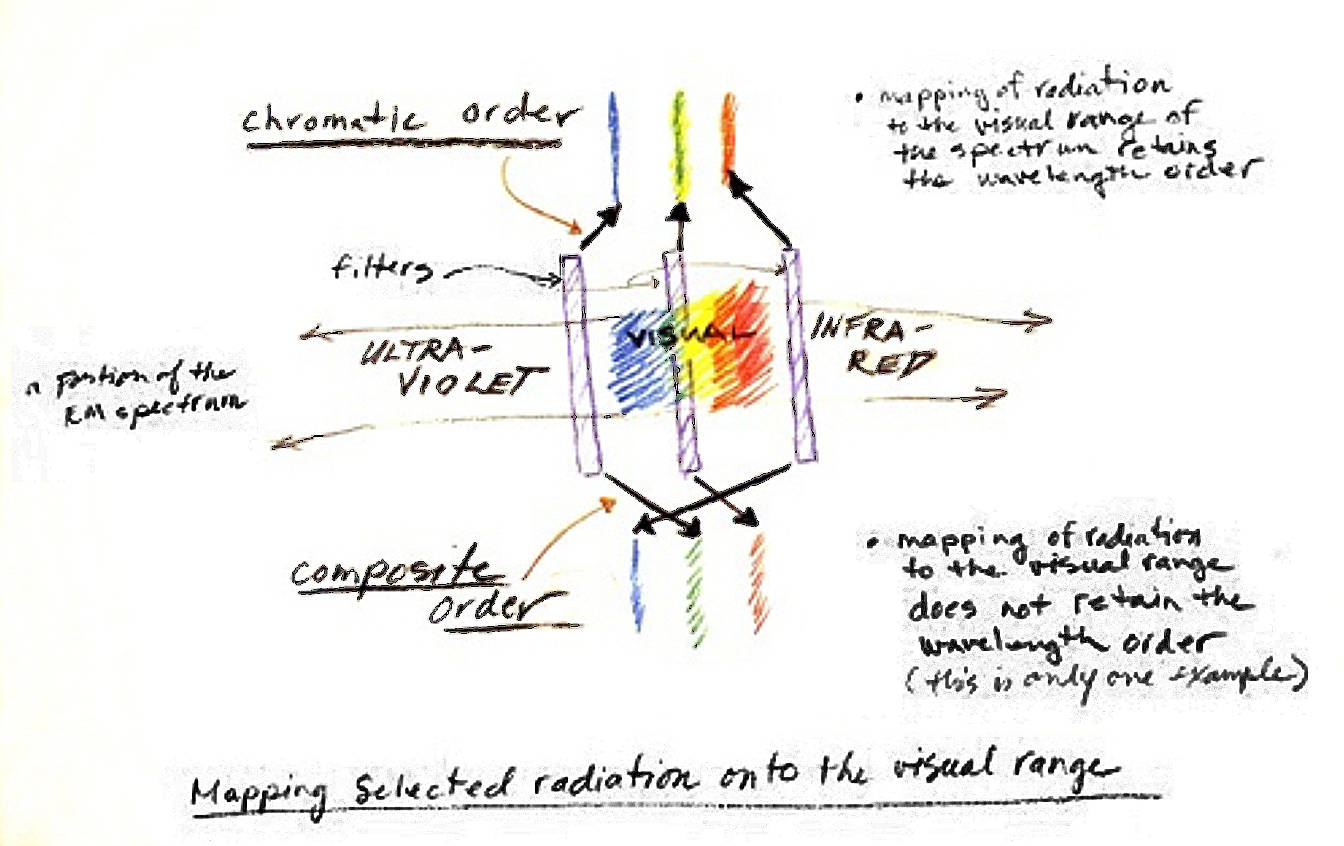,width=3.65in}} 
\caption{A photograph of the original diagram sketched by the author and used by Hubble Heritage for defining the nomenclature for colour assignments.  \label{figchromcomp}}
\end{figure}

The Hubble Heritage Team uses the nomenclature in Fig~\ref{figchromcomp}. If the data image from the filter which samples the highest energy is assigned the colour blue and other filters are  assigned colours progressively through to red as their selected energies  
decrease, this ordering is called ``chromatic". The famous Hester \& Scowen image of the Eagle Nebula\cite{EaglePillars} orders filters in this mode, which has been described as both ``natural'' and ``enhanced", since the colours assigned to individual filters do not match human perception at those energies.\footnote{When blue, green, and red are applied to narrowband filters acquiring [O III], H$_\alpha$, and [S II] emission respectively, amateur astronomers call this scheme the Hubble Palette.} This ordering suits the three images in the top row of Fig.~\ref{fighodge301} that are acquired through broadband filters. 

However,  what about the bottom three narrow-band images in Fig.~\ref{fighodge301}? 
These fall within the visual perception locus in Fig.~\ref{figfilterCIE} and we could assign colours according to each filter's central wavelength.  However,  there is a difference in texture and structure between the hydrogen emission in the 656 nm filter and the sulphur emission in the 673 nm filter, while both of them are perceived as red. We would lose the information that there are different elements, behaving differently, in this nebula if we assigned colour according to what our eye-brain system would see. 

There is a different option, labelled ``composite" in Fig~\ref{figchromcomp}.  This colour assignment is not to be confused with the final image being a ``compound'' of several visual elements.  Here ``composite'' refers to colour assignments that are not ordered in a monotonic fashion corresponding to energy decrease in the EM  spectrum.  This does not mean that the ordering is arbitrary or random. Indeed the colours can be selected to create a distinction that highlights the science, creates spatial depth and a richness of detail, thus engaging the viewer's attention. 

\subsection{Selecting harmonious colours, colour wheels and working with the human mind.}\label{colourwheel}

Colour is a qualia, meaning that the perception of colour arises from the stimulation of our eyes by phenomena and is not a single physical entity in and of itself.  As demonstrated by the after-image in  Fig.~\ref{figafterimage}, colour is generated by the eye-brain system and not by EM spectral wavelengths alone. The physiology that produces colour\cite{eye} is 
beyond the scope of this paper.  However,  the effects generated by physiology and the colour relationships, observed  by visual artists and image-makers,  are relevant to the construction of a striking image. 

\begin{figure}[]
\centerline{\psfig{file=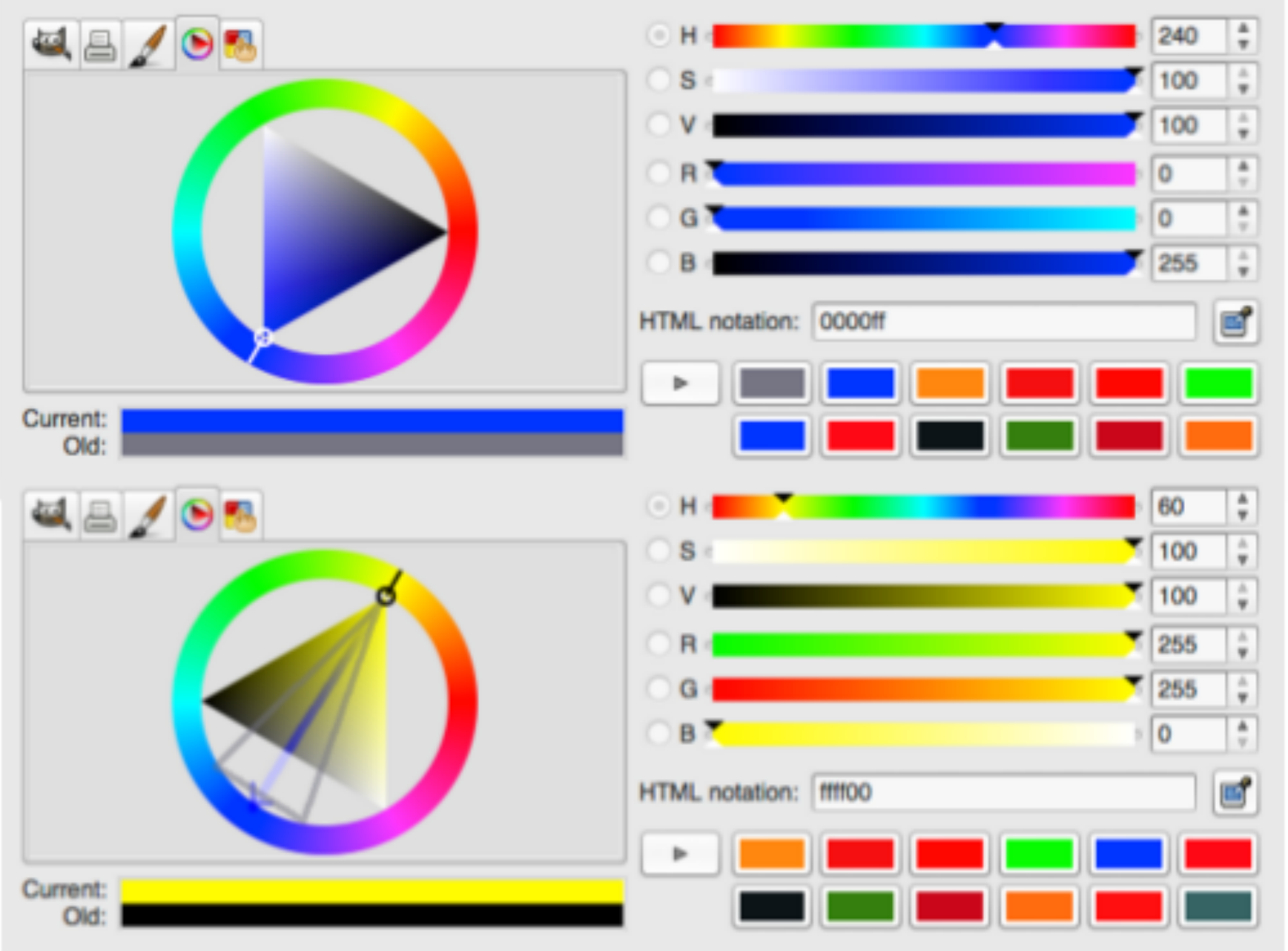,width=3.25in}} 
\caption{Colour wheel tool provided by GIMP\cite{GIMP}.   Top:  The 3 primary colours in the additive process are at the apexes of the central triangle. Bottom: Directly across the wheel from yellow is its complementary colour B, as indicated by the purple arrow.  The complement can be split into 2 colours equidistant from B along the wheel.  One example is given by the grey triangle. Another is the triangle with equilateral sides and apexes at cyan and magenta. (One should take care when using these software tools.  While the  dialog box shows useful RGB values, the Hue, Saturation, Value colour space (HSV) is not true to colour perception.  Additionally all the hues in this wheel are not the same brightness, as they would be in an accurate colour wheel.)  \label{figcolourwheel}}
\end{figure}

Fig.~\ref{figcolourwheel} shows  a  colour wheel from the GIMP\cite{GIMP}, which has an ancestry that includes  Isaac Newton, James Clerk Maxwell, Thomas Young, Johann Wolfgang von Goethe, Moses Harris, and Johannes Itten\cite{itten}.   It is a handy tool for 
understanding the relationships between colours.  Individual colours are created by combining the so-called ``primary colours'',  which are defined according to whether the mixing process is additive or subtractive. 
When creating images on a monitor the process is additive, 
which uses ``red", ``green" and ``blue", as its primary colours -- familiar as RGB.  
The RGB colour guns in a monitor have wavelengths of roughly 460 nm,   550 nm and 605 nm, each with a spread of at least 10 nm that depends on the monitor's hardware and on the portion of the human CIE colour space 
that it can reproduce.\footnote{Hardware, paint, printing inks, etc.~cannot match all hues detected by the eye-brain system. The complete set of colours that can be produced is called a ``gamut'' and these can be drawn as regions in the CIE diagram. An example is sRGB, a standard gamut used by Microsoft and HP for images that will be displayed on the internet.}  Thus,  interestingly, R and B actually fall on the CIE perception locus in Fig.~\ref{figfilterCIE} in the realms of  orange-red and purple-blue respectively.  Notice on the colour wheel that these primary colours are at the apexes of an equilateral triangle.   

Midway between the primaries on the colour wheel are the so-called ``secondary colours''. These are produced by mixing two primaries. 
The secondary colours in the additive process, which are cyan (C), magenta (M) and yellow (Y), are the primary colours in the subtractive process, which applies to paints and printing inks. (Similarly the secondary colours in the subtractive process are the primary colours in the additive process.)   
The term ``harmonious" in visual art is not an opinion but rather a definition based on physiology and can be described using the colour wheel.  In the additive process, combining the three primary colours together in the correct proportion will result in pure white being perceived. This result defines ``harmonious". A harmonious combination of hues in the subtractive process produces black or neutral dark grey, without a suggestion of colour. If the proportion is not quite right then a colour that is not harmonious, such as brown, results.

Directly across the colour wheel from an individual colour 
lies what is termed its ``complementary" colour.  Let us use yellow, Y, as an example.  Across the wheel from Y is purple-blue (B). 
Notice that the pair 
consists of all three primary colours  since R + G = Y. 
Thus combining the complements can generate white -- so they are harmonious.   Also notice that B is the colour one sees as an afterimage if one stares at Y for 15 seconds (\S~\ref{perception})!   This production of the complement is based on our eye-brain response to factors such as reflectance, luminance and wavelength associated with viewing the initial 
colour.  Thus to create a striking, harmonious image one can select complementary colour pairs in order to support, and resonate with, this physiological response. 
 
Since we typically construct public outreach images out of  a minimum of three datasets we need to select our input colours so that more than two complementary colours appear in our resultant image.   One strategy is to ``split" the complement.   Again selecting Y as one of the colours, instead of using B, one can select colours to either side of B on the colour wheel --- they should be an equal number of degrees away from B.  For example this split complement would consist of two hues in the blue family,  one with more purple in it than B and the other with less purple than B (Fig.~\ref{figcolourwheel}, bottom).  Note that a triangle with two equal sides and a shorter base can be formed in the colour wheel with these colours at the apexes. Also, if all sides of the triangle are equal then, for Y,  the apexes are the three secondary colours (CMY). Hues selected with this equilateral triangle recipe can be used to produce the strongest colour  contrast (pure hue), which we discuss in \S~\ref{contrasts}.    Other strategies for creating harmonious images are to select colours from the colour wheel using rectangles and squares\cite{itten}. 


\subsection{Composition: How is your eye led in a picture?}\label{composition}

A photograph or painting placed on a wall presents the viewer with a  2-D rectangle referred to as a ``picture plane" and the motion of the eye across this picture plane, i.\/e.~ how it is read, appears to be 
similar in most cultures. That is, the visual reading of an image is independent of the direction that a culture reads text in its language. In art college we are informed that our progression through the picture plane starts at about 1/5th of the way up the left edge and in a few seconds zips out the top right corner. If there are no elements in the image to obstruct this progress the viewer is not engaged with the picture; that is, they feel no need to continue to look at it and it is forgotten.   Artists discovered how to retain the viewer's attention using composition, which is the arrangement of elements to guide the eye.   They will block the eye's rapid progress out of the picture plane by using vertical, horizontal and diagonal elements.    
The painter Piet Mondrian, as an example, produces a trajectory using blocks and rectangles of only three colours plus black and white. 
While to the general viewer Mondrian's paintings may appear to be child's play, 
 an expert who is used to monitoring the motion of their own eye can detect forgeries\cite{JansonMondrian}.  

\begin{figure}[]
\centerline{\psfig{file=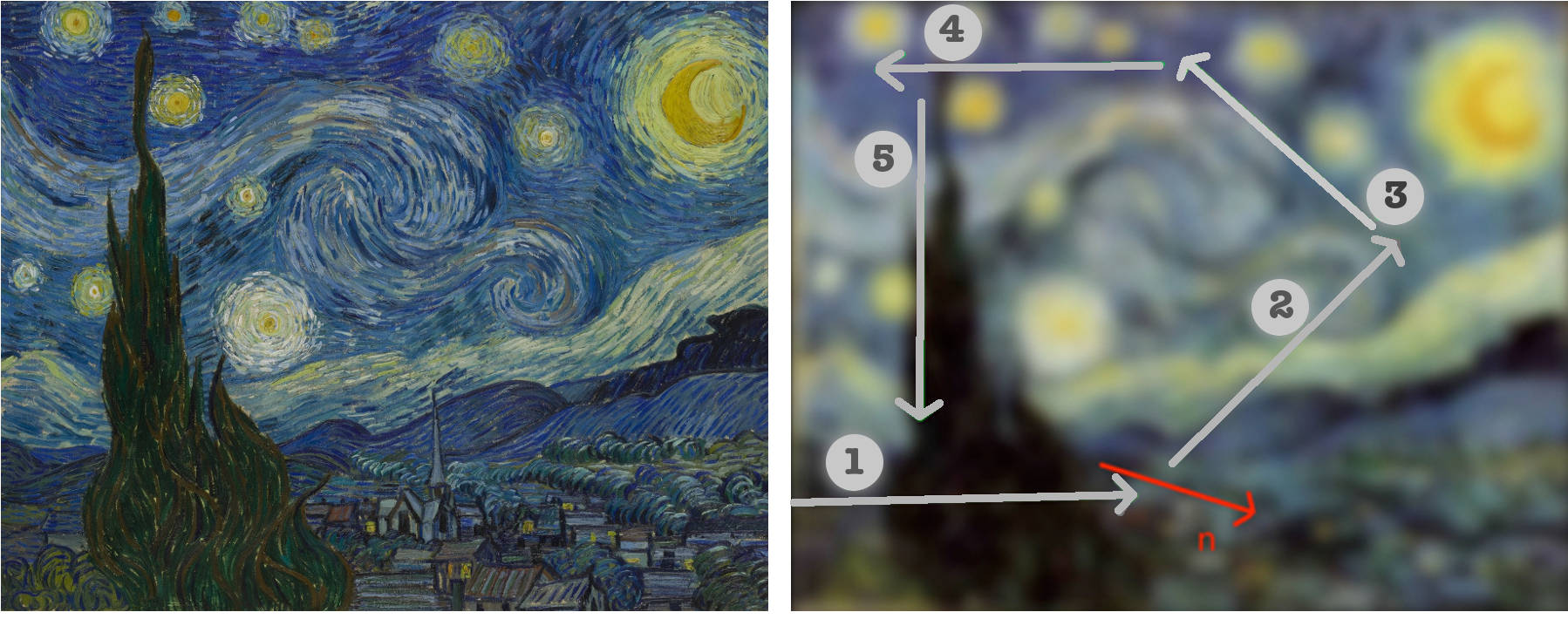, width=5.0in}} 
\caption{Composition and The Starry Night. This painting by Vincent Van Gogh, owned by the Museum of Modern Art\cite{starrynightMOMA}, contains elements (or structures) to impede the eye's exit from the picture plane. The schematic on the right shows some of the paths that the eye follows and the one labelled ``3" 
traces a ``virtual diagonal". 
The final (nth) path leads to the church, which is not in the static centre of the picture plane.   \label{figVGstarrynight}}
\end{figure}

I typically use Fig.~\ref{figVGstarrynight}  as an example of how artists guide the eye.  In ``The Starry Night"\cite{starrynightMOMA},  Vincent Van Gogh slows the progress of the viewer's eye, as soon as it enters the bottom left edge of the picture plane, by placing a dark vertical mass of trees on the left. As the eye now moves more slowly towards the top right corner of the tableau, pulled by the bright moon,  the viewer encounters a swirl of clouds, the upper edge of which trace a diagonal - I'll refer to this as a ``virtual diagonal" and it is traced with a white line (\#3) in the right-hand diagram.  Notice  the shepherding stars that help lead the eye back to the left toward a bright star in the upper corner. From there the eye slides down the vertical mass of trees towards the point of entry and starts to make another lap through the painting on a slightly smaller orbit. A number of contracting laps are traced until the eye apprehends the church in bottom of the painting. Although this has taken some time to describe, this visual circuit has been completed in seconds. 

Also notice that  
the church, is not in the dead centre of the painting.   That would be as static, and as boringly comprehensible, as a bull's eye target -- 
the viewer would instantly assess the subject and 
stop looking at the picture plane. The alternative  off-centre, dynamic placement retains the viewer's attention. 


\begin{figure}[]
\centerline{\psfig{file=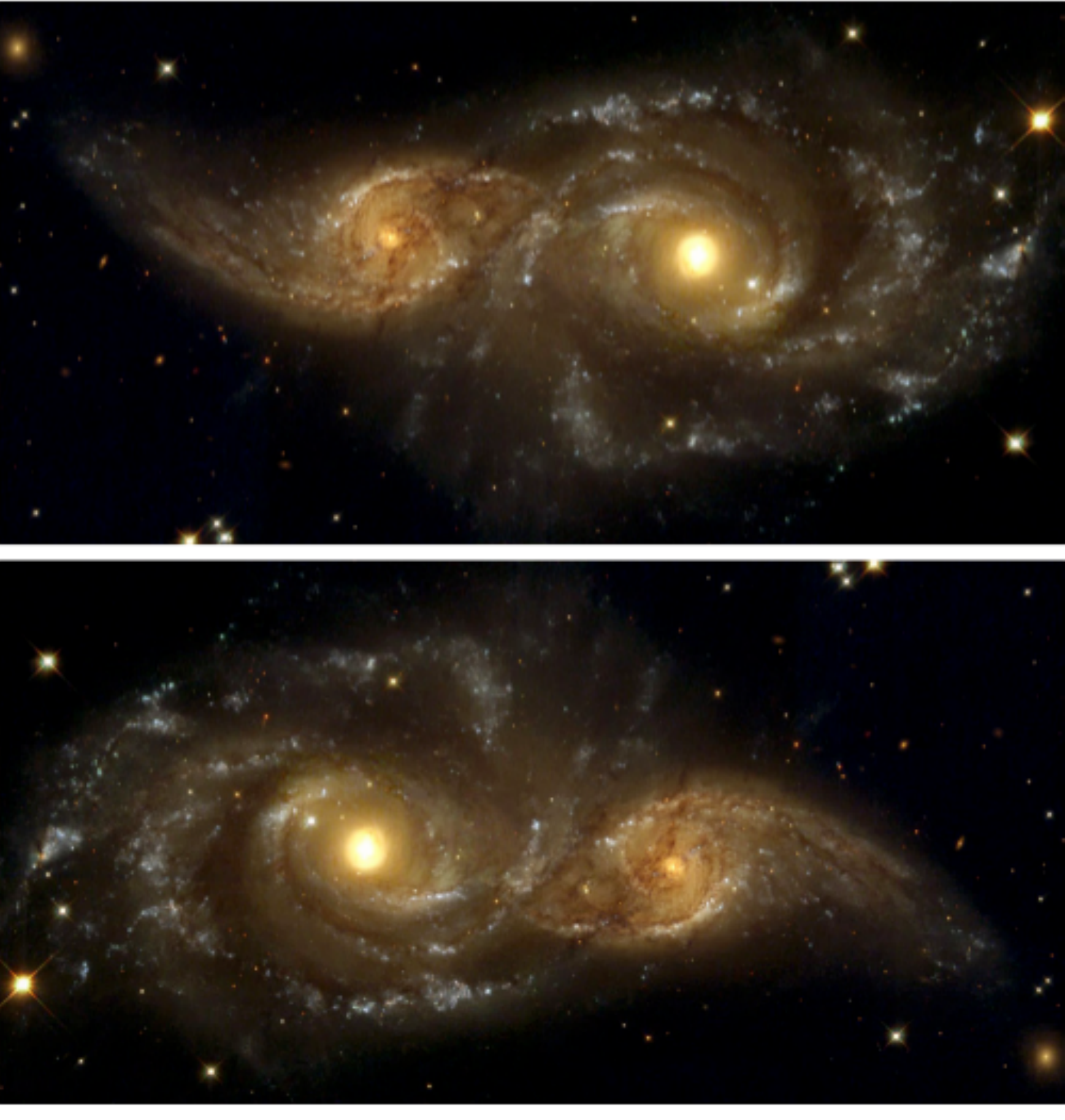, width=4.0in}} 
\caption{Composition and NGC 2207 and IC 2163. The top image  has the 
conventional (east to the left) orientation for astronomical maps.
The image has little depth. However,  a spiral arm appears to swing towards the viewer in the bottom image, which is simply the top one rotated by 180 degrees. In the bottom orientation, used in the Hubble Heritage release,  the galaxy disks align to form a virtual diagonal as in Fig.~\ref{figVGstarrynight}. 
This image appeared in numerous popular magazines in the bottom orientation and  in a professional article\cite{ngc2207pubs} in the top orientation. \label{figNGC2207}}
\end{figure}

Fig.~\ref{figNGC2207}, which compares two orientations, provides an astronomy example.  The convention of presenting sky maps with north to the top and east to the left facilitates comparing them with the constellations when the top of one's head is pointed at the north pole.   However,  the Hubble Heritage\cite{ngc2207pubs} image of galaxy NGC~2207 and its interacting companion IC 2163 appears as flat as two eggs on a cast iron frying pan in this orientation (top image).   A simple rotation of 180 degrees and a tight cropping (but no cutting and pasting), sets up a composition similar to ``The Starry Night" (bottom image).    Fortuitously there is a bright star where the eye enters the picture plane and one in the top left corner.   So when the eye encounters the virtual diagonal barrier formed by the swirl of the upper edges of the galaxy disks, it is drawn towards the upper left as in The Starry Night.   Then it is drawn back down to the bright star at the entry point, and starts another lap in the image.   This initial engagement allows the eye-brain system to experience the more resolved detail in the spiral arms at the bottom of NGC~2207, causing those to pop forward towards the viewer.  Thus this orientation creates depth and the subject is of more interest than in the ``flat frying pan'' version.  Indeed when 
when this comparison  is made during public talks some people do not believe they are looking at the same picture in two different orientations and ask me to blink between 
orientations. The Hubble Heritage Project's ``upside down" (bottom) 
configuration was 
in Life Magazine's ``The Year in Pictures" 1999 issue,  attesting to its success as an engaging image. 

\subsection{The Reading of Colour: Which is hot and which  is cold? }\label{readcolour}

\begin{figure}[]
\centerline{\psfig{file=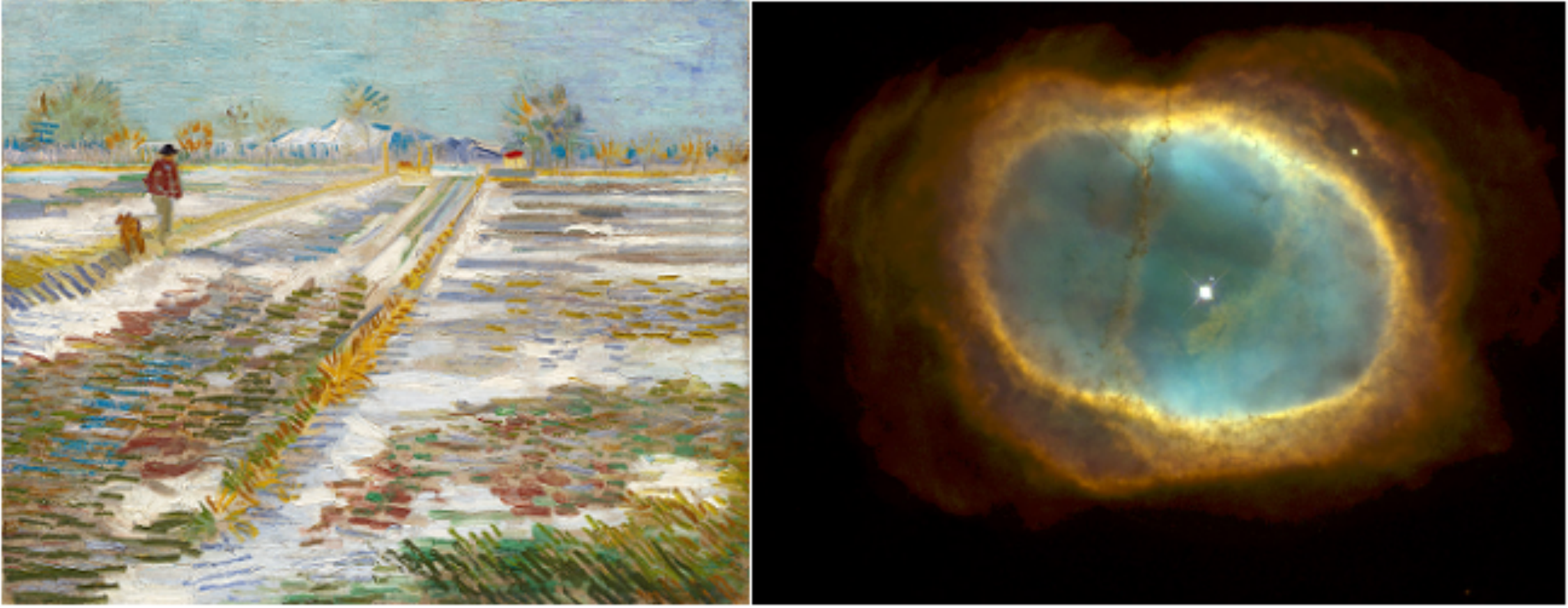, width=5.0in}} 
\caption{Warm, Cold, and Snowy Landscape versus a Planetary Nebula. Left: ``Landscape with Snow'' by Vincent Van Gogh, owned by the Guggenheim Museum\cite{VGlandGug}, is one of many examples where snow and ice, which have a cold temperature,  are colored blue. Contrast this with scientists' use of blue to represent high energy,  as in the blue colour of the hottest part of a candle flame. 
Right:  Planetary Nebula NGC~3132\cite{HHPN3132}.  Constructed by the Hubble Heritage Team, this image uses a ``warm" blue, that is one that will more likely appear closer to the viewer than primary B  (\S~\ref{perception}). Nevertheless the general public is likely to read the ionized central region as cold. \label{figVGlandPNe}}
\end{figure}

The van Gogh landscape\cite{VGlandGug}  in Fig.~\ref{figVGlandPNe} also guides the eye, with strong diagonals and a horizontal horizon.  
Now focus on the colour in this image.  We can see that the cold snow and ice are blue while the warmer mud is reddish.  Blue is generally read as cool \footnote{e.\/g.~Lauren Harris's ``Snow Fantasy", Tom Thompson's ``Snow Shadows", Claude Monet's ``Stack of Wheat" and ``Snow Effect" and a myriad of others.} while reds and yellows are considered warm \footnote{e.\/g.~J.~M.~W.~Turner's ``The Burning of the Houses of Lords and Commons"; 
paintings of fire.}.  If this is the case, how does a non-science audience comprehend the temperatures of planetary nebula NGC~3132 in Fig.~\ref{figVGlandPNe}?  A showing of hands at my lectures demonstrates that the majority read the cyan as  cool and the dusty red edges as warm.  

Since this image was constructed by the Hubble Heritage Team, consisting of scientists,  the colour scheme follows that of excitation energy.   That is, lower energy electromagnetic radiation is assigned red while blue represents hotter radiation.     The philosophy in this case was to produce a pedagogical tool to inform the public about the colours associated with energy ranges of light.  The supplemental online material\cite{HHPN3132} draws an analogy with the temperature ranges in Yellowstone's Grand Prismatic Spring, which for biological reasons is blue at the highest temperatures. 

This example highlights 
 that {\it more than one colour rendition of astronomical data is valid}.   Firstly, the assigned colours do not match what the eye would see; H$_{\alpha}$ is assigned green, not red.  So other image-makers need not be committed to Hubble Heritage's  ``untrue to the eye", yet illuminating, colour scheme. Secondly, while a  concession was made to the 
usual reading, 
in that cyan is a warmer version of blue 
than ultramarine, one could alternatively create an image that eliminates the need for a legend when presenting this object to the non-science public.  That is,  reds and yellows -- referred to by artists as {\it warm colours} -- could be used in the centre, while blues and greens -- so-called {\it cool colours} -- could be used in the outer region. 

\subsection{Creating Spatial Depth with Colour: Which colours stand out and which recede?}\label{depth} 

\begin{figure}[]
\centerline{\psfig{file=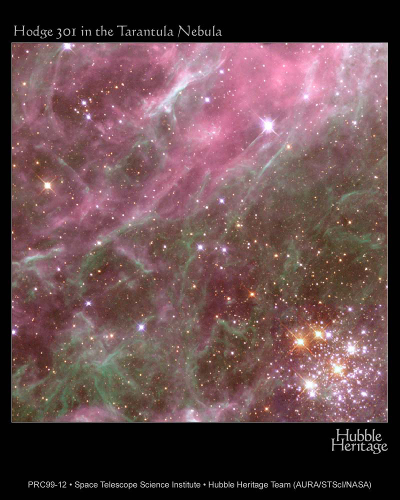, width=2.5in}} 
\caption{Depth, Contrast and Hodge 301. The 6 data sets in Fig.~\ref{fighodge301} have been assigned colours, both in chromatic order and composite order (\S~\ref{motive}), then combined. As well as creating a sense of depth, complementary colours resulted, making this an example of complementary contrast (\S~\ref{contrasts}). \label{fighodge301HH}}
\end{figure}

Returning to van Gogh's ``Landscape in Snow" (Fig.~\ref{figVGlandPNe}) note that the reddish and yellow colours are dominant in the region where the eye enters the picture plane, that is, in the foreground,  while the background is literally sky blue.  A property of warm colours is that they come forward while cool colours sink back; as   demonstrated in Fig.~\ref{figwarmcoolballs}.    This effect can be used to create spatial depth in a picture.

Returning to our dilemma about how to colourize the three narrow-band filters in Fig.~\ref{fighodge301}, we can now see a motivation for using composite ordering for colourizing the data (\S~\ref{motive}).   If we chose relatively cooler colours for the gas in the middle section of the nebula we can help make the star cluster sit forward in our composition. The final rendition, Fig.~\ref{fighodge301HH} 
 includes the complementary colours magenta and green, with the cooler of those two colours (green) filling a broad region near the cluster and the warmer colour assigned to the gas spatially further away from the cluster.    

\begin{figure}[]
\centerline{\psfig{file=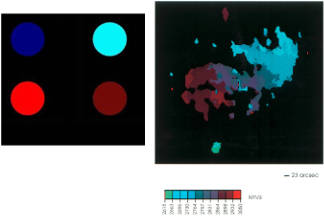}} 
\caption{Depth and Science Figures. Left: a comparison of the circles from Fig.~\ref{figwarmcoolballs} with an alternative scheme for ``red''  and ``blue''. The cyan blue is visually warm and perceptually appears to come towards the viewer.  This is enhanced by the light/dark contrast between the ball and the background.  Similarly the darker shade of red visually recedes since there is less contrast.  Right:  Applying the alternative scheme to a 21 cm  velocity field map of NGC~3256\cite{NGC3256}. This figure retains the convention of assigning ``blue" to blue-shifted emission from gas which is approaching. Yet  the torquoise and cyan colours perceptually support the definition of blue-shift.The viewer can visually apprehend the  line-of-sight trajectories (Eq.~\ref{dopplerv}) of independent clouds as well as the rotation of this late stage merger system of two galaxies. 
 \label{figvelfield}}
\end{figure}

The left side of 
Fig.~\ref{figvelfield} demonstrates two more effects.  Firstly, the colour or darkness of a surrounding region can effect our perception.  Secondly, one need not give up a conventional colour legend -- rather one can manipulate the colours such that the visuals support the intended information.   Note that the right ``blue'' circle is cyan - a warmer blue (i.\/e.~towards yellow in the colour wheel) than the blue in the left circle.  Note how the cyan pops forward relative to the dark red, which is almost co-planar with its dark surround.  (See Itten\cite{itten} for other effects.) 

This knowledge can be helpful for rendering scientific information in both outreach images (\S~\ref{scimean}) and research illustrations. 
Fig.~\ref{figwarmcoolballs}  demonstrates the convention for representing the Doppler red-shifted receding gas and blue-shifted approaching gas  are assigned colours that perceptually cause the receding gas to pop forward and the approaching gas to sink towards the background. Contrary to this, the cyan-to-dark-red colour scheme applied in the right-hand side of Fig.~\ref{figvelfield} causes blue-shifted emission to pop to the foreground,  resonating with the intended information. 
This visual scheme supports our knowledge of not only how the galaxy is rotating (i.\/e.~right side approaches as left side recedes) but also highlights the motion of nearby clouds of gas relative to the galaxy. 

\subsection{Colour Contrasts: Visually shout or whisper.}\label{contrasts}
\begin{figure}[]
\centerline{\psfig{file=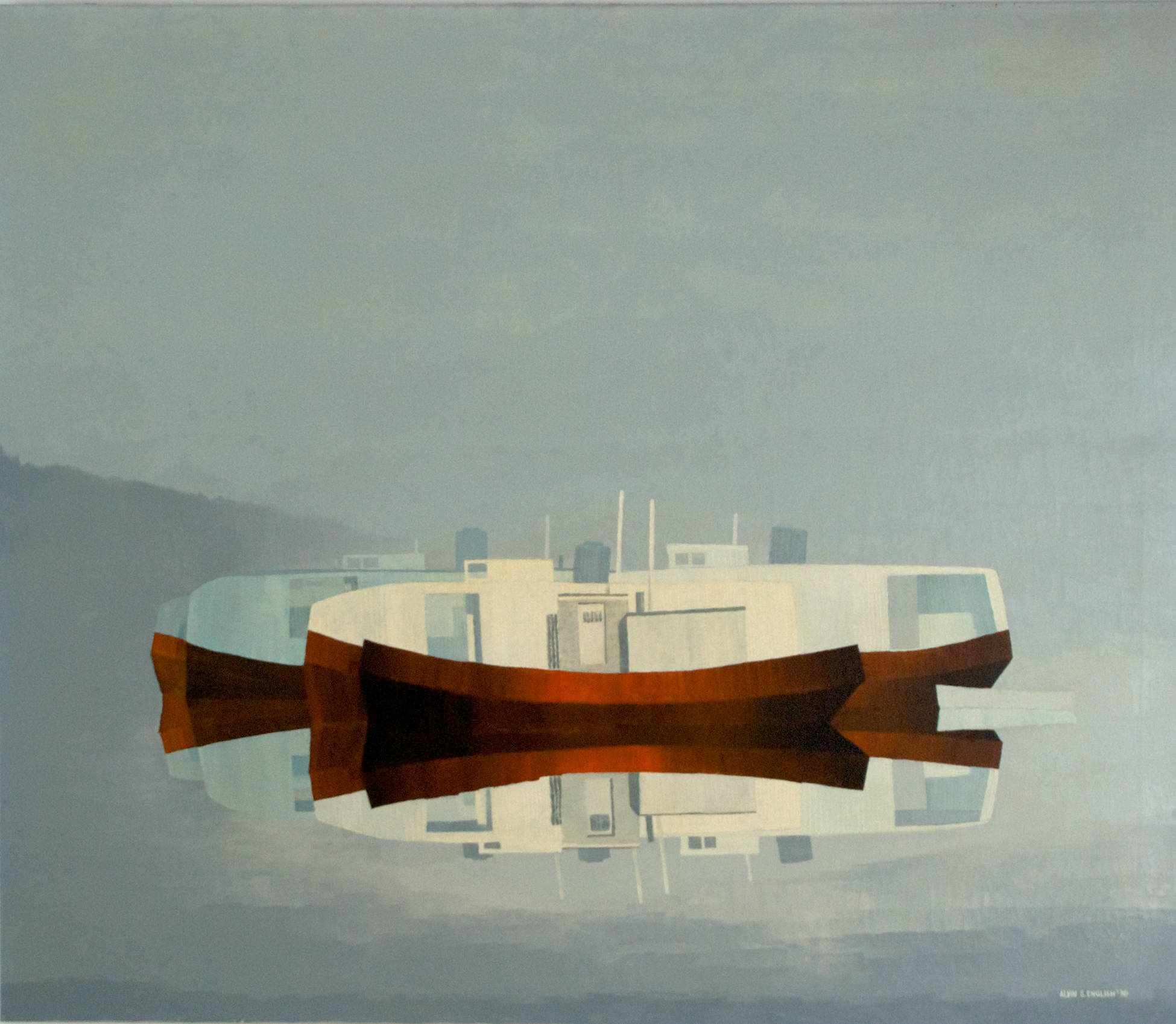, width=3.5in}} 
\caption{Multiple contrasts. ``The Drift'', by Alvin S.~English, predominantly  exhibits the contrast of saturation (Fig.~\ref{figltdrksat}) in which both the purity of colour is varied and the intensity values are juxtaposed, in this case subtely.  
The distinction between elements is achieved using orange and blue
which also  provide two other contrasts --- 
 complementary  and cold-warm. 
\label{figAlvinEnglish}}
\end{figure}

Finally we describe the contrasts that enhance the impact of colour in imagery.  In black and white images there is only one kind of contrast -- light-dark.   For this mode of creating distinctions between visual elements the extreme example is simply white against black.  However,   extreme variations in greyscale values need not play a dominant role when colour is also used to create distinctions, as in ``The Drift" in Fig~\ref{figAlvinEnglish}.  Indeed Itten\cite{itten} describes seven contrasts, using a geometric abstract diagram for each.  This section presents Itten's contrasts  in the six figures below that are constructed using the same HST data of a portion of the Orion Nebula\cite{heritageOrion}.  For each figure, colour has been assigned, as  in \S~\ref{construction}, steps \ref{layers} and \ref{colourize}, to each of four different filters: B, V, z, and H$_\alpha$.  
(The first three filters are very similar to broadband b, g, z  filters on the left-hand side of Fig.~\ref{figfilterCIE}; the last is a narrowband filter described in the caption of Fig.~\ref{figMalEngCrab}.) The combined layers were adjusted as in step \ref{combine} to attempt to make one particular colour contrast dominate per figure.  

\begin{figure}[h!]
\centerline{\psfig{file=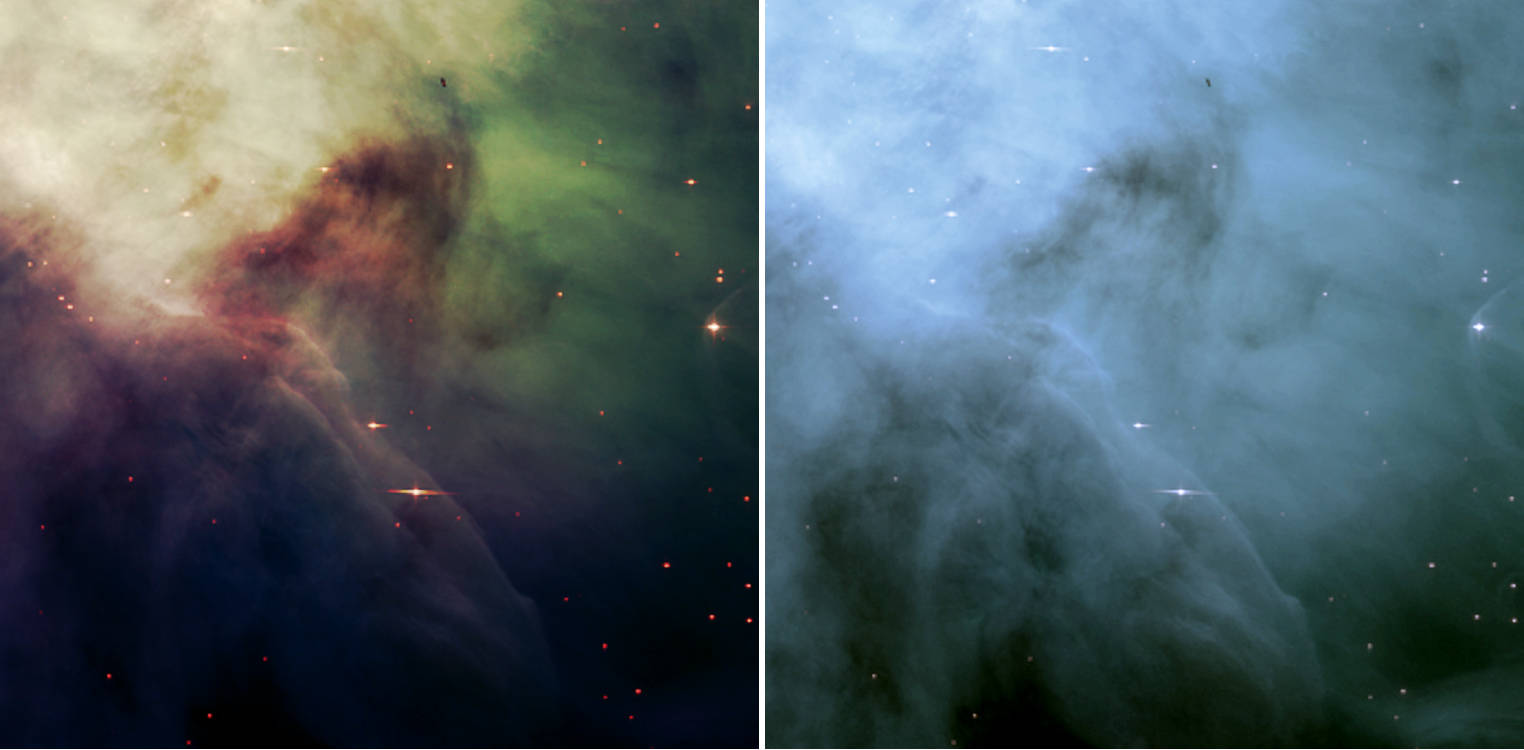, width=3.4in}}  
\caption{Light-dark and saturation contrasts.  These are appropriate for colour blind viewers. Left: Light-dark contrast juxtaposes brightnesses. Right: Saturation contrast involves surrounding  more pure colours with dull, less pure colours. The decrease in saturation in the latter colours can be accomplished by adding either white, black, grey or complementary colours. (Orion Nebula.)}
 \label{figltdrksat}
\end{figure}

If the image-maker wishes to be inclusive with respect to colour-blind viewers, then light-dark or saturation contrasts, as in Fig~\ref{figAlvinEnglish} and Fig.~\ref{figltdrksat},
are appropriate.

\begin{figure}[h!]
\centerline{\psfig{file=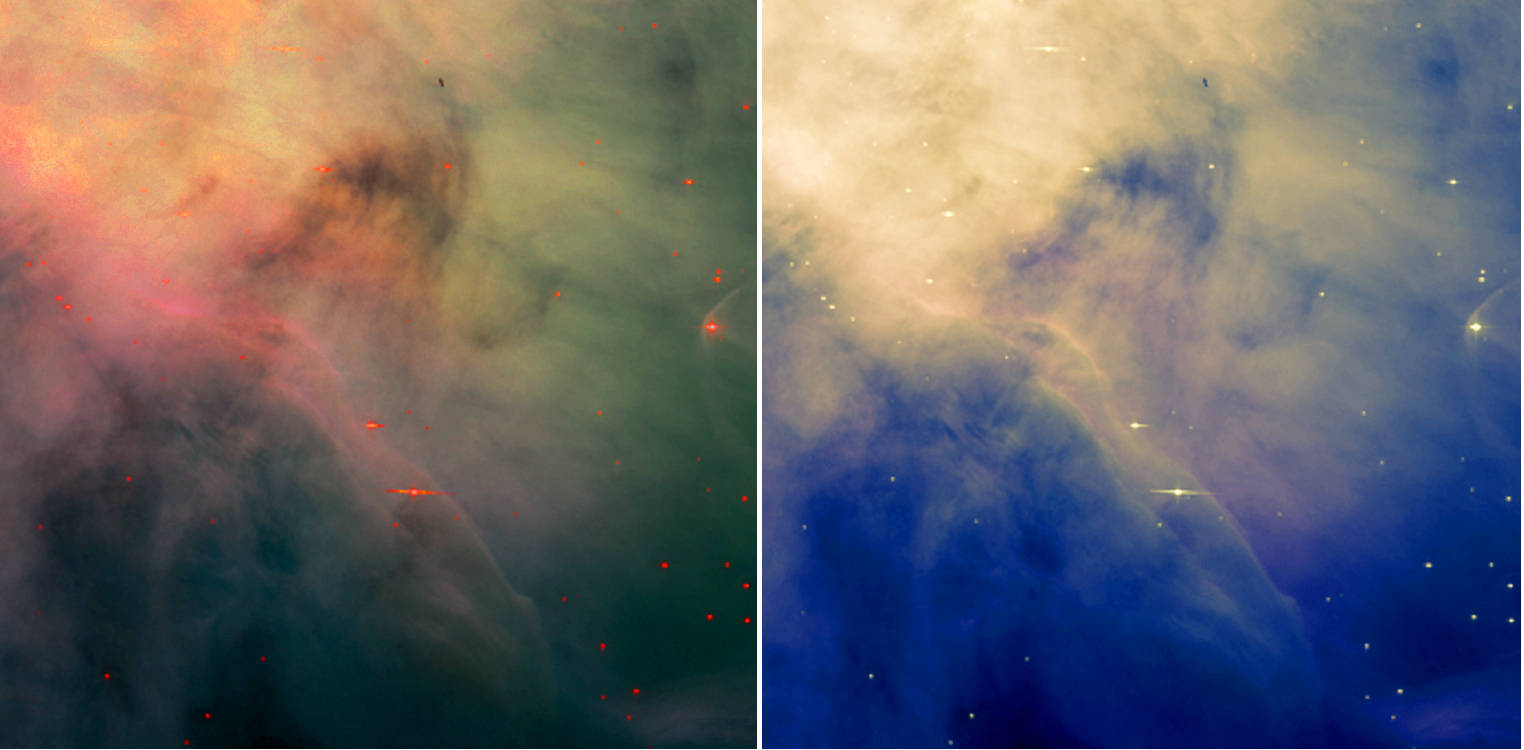, width=3.4in}}  
\caption{Hue and cold-warm contrasts.   Left: Hue contrast juxtaposes saturated colours, including black and white.  The colours can be randomly selected.  Right: Cold-warm contrast juxtaposes cool colours, such as blue, with warm ones, such as red or yellow. Also see Fig.~\ref{figwarmcoolballs}. (Orion Nebula.)}
 \label{fighuecool}
\end{figure}

Hue contrast, which uses pure saturated colours, is so ``loud'' that it is popular with advertisers for grabbing a reluctant audience's attention. In the science context, it is typical in contour plots as well as images like the left-hand side of Fig.~\ref{figCrabBlairHST} and those produced by the Spitzer  Space Telscope and Chandra X-ray Observatory imagemakers\cite{spitzerchandra}. Cold-warm, another strong contrast, is a common result for novice astronomy image-makers and helps create spatial depth (\S~\ref{perception}). See Fig.~\ref{fighuecool}.

\begin{figure}[h!]
\centerline{\psfig{file=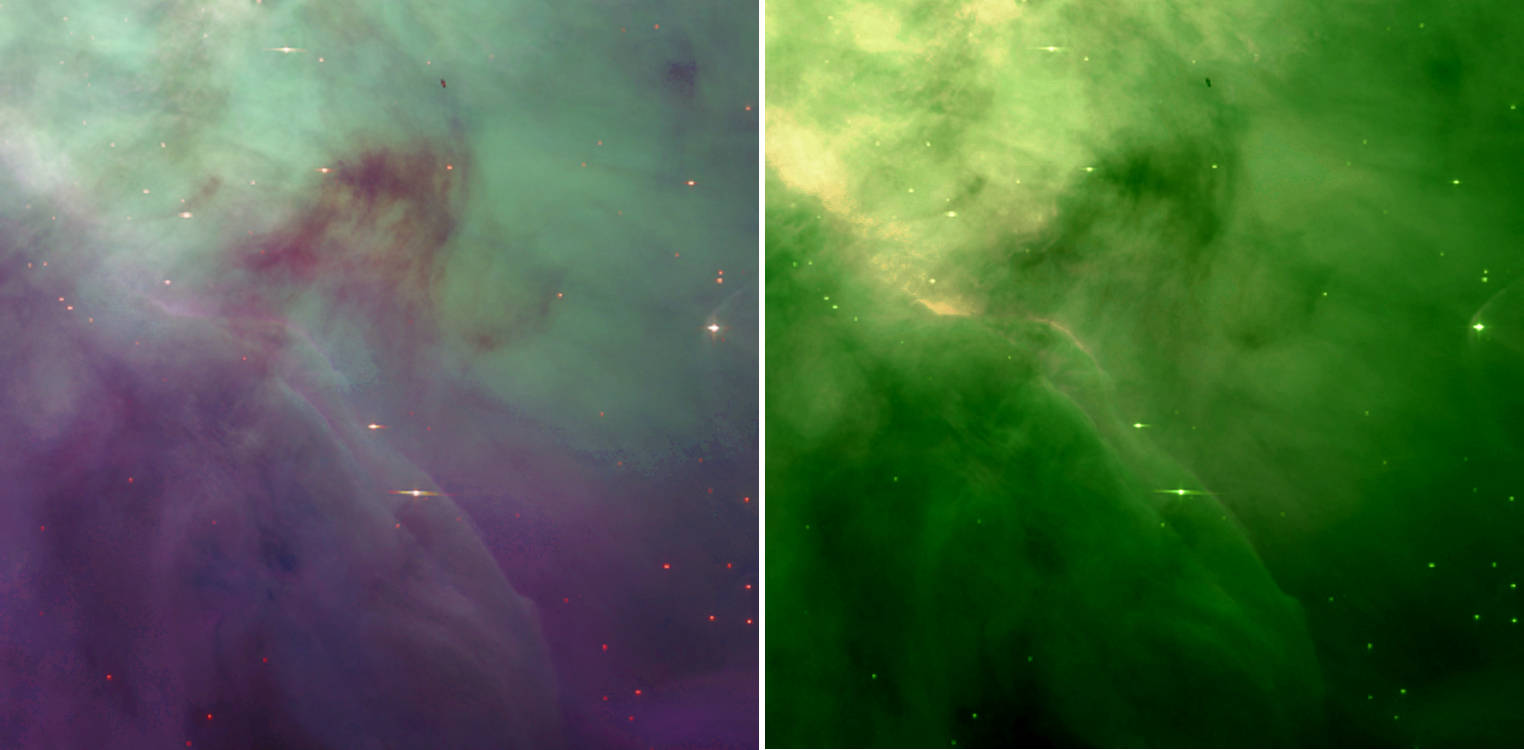, width=3.4in}}  
\caption{Complementary and extention contrasts.   Left: Complementary contrast juxtaposes colours across from each other on a colour wheel (Fig.~\ref{figcolourwheel}).  Multiple colours can be selected by ``splitting" the complements (see \S~\ref{colourwheel}).  Right: Extension contrast juxtaposes differently sized areas to achieve balance. Here a pink 
thread of emission is balanced by green 
 areas. (Orion Nebula.)}
 \label{figcompextent}
\end{figure}

Images that use complementary contrast, including splitting the complement, e.g. right-hand side of Fig.~\ref{figCrabBlairHST}, are more complex. This and simultaneous contrast, which  uses visual physiological effects like that in Fig.~\ref{figafterimage} to create luminous colours, were used by the Impressionists, Pointillists, as well as van Gogh and others. The effects created by simultaneous contrast rely on the size of points, since colour mixing is to occur in the eye-brain system (rather than occurring on a canvas via the blending of paint). Also  a role can be played by the size of the area that is intended to glow and the size of the surrounding colour region.  Therefore simultaneous contrast is rare in astronomy images since image-makers are constrained by the spatial scales of the morphological features in their data. Since the Orion Nebula data do not lend themselves to this contrast, a figure is not provided. Instead in Fig.~\ref{figcompextent}  the contrast of extension is presented along with complementary contrast. 

Artists intentionally use more than one colour contrast in a single image (e.\/g.~Fig~\ref{figAlvinEnglish}). 
Usually the process of combining coloured data layers when constructing an astronomy image (\S~\ref{construction}) produces more than one contrast.   


 \ \\
To summarize \S~\ref{visart}, there are many valid combinations of orientations, croppings, colour harmonies, contrast schemes, goals and intentions that retain a viewer's attention and can be used to highlight salient scientific features.  While  the astronomy image-maker  needs to experiment, observing what combinations work for a particular target as well as for their scientific message, many different combinations of these elements generate  equally powerful compositions that will  resonate with a viewer's physiological response to visual stimuli.

\section{Retaining Scientific Meaning}\label{scimean}
\subsection{Map or Spacescape?}\label{manipulation}
Using visual grammar (\S~\ref{visart}) provides the means for converting data, i.\/e.~logic tradition maps, into convincing western image tradition depictions of phenomena (\S~\ref{construction}).  A scientist may fear that the use of composition and colour to create images that retain the viewer's attention may obliterate the information inherent in the data while the public may fear that the image is a cut-and-paste ``spacescape" fantasy.   The astronomer image-maker attempts to avoid both of these situations.  

Instead of creating an objet d'art, the astronomer image-maker's goals are  to illustrate  
the scientific understanding of the moment.  This constrains the image manipulation process and the resultant pictures.  When astronomers present a compact group of galaxies with intriguing tidal features they want it read as a physically 
intriguing astronomical object, not a creative montage of non-interacting galaxies.  Similarly they avoid the over-application of algorithms, such as ``sharpening'', which create artefacts that could be construed as real features by a naive viewer. Zoltan Levay, an image-making leader, restrains the manipulation of HST images to software approaches that are analogous to traditional film darkroom techniques,  
e.\/g.~dodging and burning\footnote{This technique involves moving an obstruction between the light,  passing through film and emanating from an enlarger, and the photographic paper. The regions which are blocked are ``dodged'' and those which have more light falling on them are ``burned''. Dark regions are burned to reveal detail while light regions are simultaneously dodged to avoid saturation, which destroys detail.} to bring out detail in bright and dark regions. 

In spite of these self-imposed limits on manipulation, outreach images are not 1-to-1 analogues with what the human eye sees.  Neither are other images created in the tradition of science culture, 
such as contour plots. Thus the outreach image-maker 
 should not be constrained to producing such a correspondence.  For one, the data do not permit a match to human vision since the filters  (or channels) are initially selected for research analysis purposes and the non-optical wavelength regimes  are often incorporated. 
 Secondly, these images are not snapshots -- the exposures can be separated by months or years and there is human intervention, via processing, manipulation and aesthetic choices.  Thirdly,  it is unreasonable to mimic the sight of the human eye since it is not a good measuring device for scientific ``truth".  Physiologically the eye does not produce intense colour when observing faint light. Thus a target through a telescope's eye-piece typically appears ghostly green 
 or grey.  For example, a planetary nebula with a spectrum  dominated by glowing red ionized hydrogen will instead appear to be green --- the eye misleads one to believe they are looking at ionized oxygen gas.   Knowing the measured 
 spectrum of the nebula, an astronomer image-maker can 
 assign red to the H$_\alpha$ filter data and balance the intensity of the layers (\S~\ref{construction}) so that red dominates and the image provides a map of the spatial location of hydrogen. 

\begin{figure}[h]
\centerline{\psfig{file=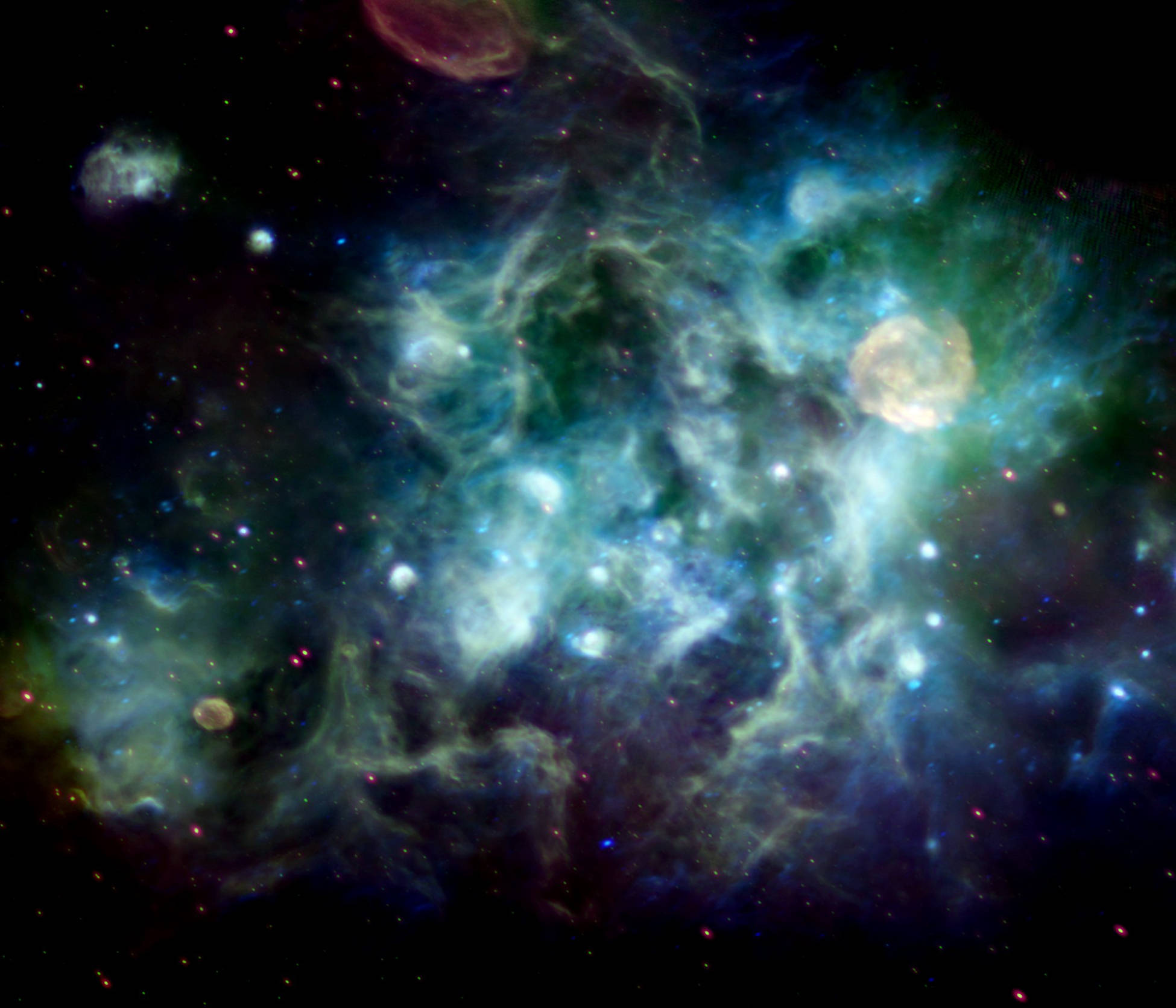, width=4.0in}} 
\caption{Recycling the Interstellar Medium in Cygnus (\S~\ref{sciexamples}).  Observations of dust were acquired by IRAS: 25 micron (blue) and data 60 micron (turquoise). Radio synchrotron emission data were acquired by DRAO: 20 cm (green) and 74 cm (rose). Interstellar plasma is green and dust is in blues and turquoise. 
Star forming clouds are white.  Supernova remnants are extended elliptical objects with yellow through red colours.  Background AGN 
have yellow cores and magenta envelopes, 
since  74 cm data have lower resolution (\S~\ref{construction}) than  20 cm data.  The cold-warm contrast (\S~\ref{contrasts}) creates depth --- the warmly coloured supernova remnants seem suspended in the foreground. (This image has appeared in popular magazines\cite{EngTayCygnus}, books, a documentary, etc.)}
 \label{figcygnus}
\end{figure}

Perhaps an analogy for outreach images is provided by the kind of Google map in which  photographs are overlaid on the substrata of street maps. When we take logic tradition maps and convert them into convincing western tradition images we attempt to make a visual object that at least resonates with the knowledge that we have gleaned from scientific analysis of the data and knowledge about the target.  

\subsection{Examples of Retaining Scientific Information}\label{sciexamples}
There are a myriad of examples of science rich images on the websites listed in \S~\ref{intropower}, including those produced by the accomplished image-makers mentioned in \S~\ref{teams} and \S~\ref{colleagueteams}. For some 
images part of their scientific content 
is apparent even 
if a legend or caption is not attached, as  figures in this paper illustrate.  

Looking at the Cygnus Region in Fig.~\ref{figcygnus} the viewer may be convinced that if they looked through a telescope's eye piece they would see this vista --- that the objects in the field of view exist. They do but this scene is not visible to the human eye.  The data were acquired in the far-infrared (IRAS) and radio (DRAO) regimes and the image is one of the first in which radio data appears to be photographic.  The colour assignments of 4 data sets are chromatically ordered (\S~\ref{motive}) such that warm IR dust components are blue and turquoise while radio continuum data  are assigned green and rose for the longest wavelength (lowest energy). The green 20 cm radiation can be either thermal or synchrotron emission.  Thus, where green is more smoothly distributed 
interstellar plasma is co-spatial with the filamentary dust structures.   The ages of elliptically-shaped supernova remnants are clearly expressed. 
Red and green produce yellow so the younger more energetic supernova remnants have this tint while the  lowest energy, oldest remnant 
 is dominated by the rose colour of the 74 cm data.   Interestingly the bottom left supernova remnant is obscured in optical data by a dust cloud called The North American Nebula --- thus this image demonstrates the discovery power of radio telescopes. All four colours add to white  so the spherical white objects are  dusty, thermal clouds that can collapse to form stars under the right conditions. Note the life cycle of star birth through star death is evident in this image. Finally there are  point-like circles with yellow centres and rose envelopes. These are also synchrotron emitting sources but not stars (which rarely emit strongly in the radio regime). These quasi-stellar objects, i.e. quasars, are the cores of background galaxies with active galactic nuclei (AGN, due to supermassive black holes) that often spew out enormous synchrotron jets. Sometimes image-makers smooth the high resolution (\S~\ref{construction}) data to match lower resolution data (such as the 74 cm observations). In this case, retaining the natural resolution allows the viewer to visually detect the AGN via their core/envelope representation. 

In the right-hand side of Fig.~\ref{figCrabBlairHST} of the Crab Nebula the ionized elements in filaments appear distinct from each other and from the more smoothly distributed optical synchrotron emission (pale blue) associated with the pulsar.  Since the elements ionize at different temperatures, a colour bar would simply map this characteristic. However,  occasionally this reading becomes complicated by the Doppler shifting of  an element's emission line due to gas motion. For example, a pink part of a filament will transition towards green as the line shifts from one narrowband filter into the wavelength range of another filter.  Also the 
gas features of Hodge 301 in Fig.~\ref{fighodge301HH} are well delineated and cool stars can be distinguished by their somewhat orange tint. The colour selections for the planetary nebula in Fig.~\ref{figVGlandPNe}  are meant to pedagogically use blue to represent ``hotter" gas residing closer to the white dwarf core (\S~\ref{readcolour}).

As described in the caption of Fig.~\ref{figCGPSplane} this image shows the sense of rotation of our Milky Way Galaxy since cooler colours representing blue-shifted gas are on the right-hand side and there is a smooth transition to the 
warm, yet dull (and therefore receding), colours representing red-shift that are on the left.  Also  purple-blue filaments occur due to an absence of emission in the channels that are assigned warm colours --- 
these filaments trace sites of HI self-absorption. Stars heat the dust (assigned pink) and the image shows that the bottom part of the excavated cone, referred to as the W4 Chimney, is thermalized by a star cluster.

The centres of the galaxies in Fig.~\ref{figNGC2207} are warm coloured since the stars are relatively cool whereas the spiral arms  of NGC 2207 are more blue since the stars there are hotter.  
Alternatively if only the HST broadband filter data are used, as in Fig.~\ref{figNGC2207},  to create Fig.~\ref{fighcg31final} then the giant ionized hydrogen regions 
appear unnaturally green\cite{releaseHCG31}.  The narrow, though strong, red H$_\alpha$ line resides in the broad wavelength range of filter centered at 606 nm that is assigned green in order to ensure the stars appear white.   However,  the gas in these H II regions, ionized by hot, UV radiating stars, are also permeated with dust.  Therefore non-optical data from two other space observatories, GALEX (UV) and Spitzer (IR), were incorporated and coloured to produce the red of ionized hydrogen without modifying the colour of most stars.

\section{Feedback into the realm of astronomy}\label{feedback}
Visual grammar (\S~\ref{visart}) can be used to construct images that simultaneously engage the attention of the  public (\S~\ref{intro}) and support the scientific content of the data (\S~\ref{scimean}). Visual representations of data and ideas are also crucial components of scientific analysis and professional publications.  To learn how to incorporate visual principles into displays of complex data one could turn to statistician and artist Edward R.~Tufte who, in revered works like ``Envisioning Information''\cite{TufteEnvision}, presents uses of colour that avoid creating visual puzzles.   Felice C.~Frankel and  Angela H.~Depace's\cite{FrankelDePace2012}   ``Visual Strategies: A practical guide to graphics for scientists and engineers'' provides a scheme for assessing visuals as well as an approach to designing clarifying figures. Images in Frankel \& Depace's {\it explanatory} category not only provide evidence of proof, as do image tradition depictions (\S~\ref{construction}), but also highlight salient points, patterns or concepts. Their category of {\it exploratory} graphics consists of images that encourage  viewers to discover information and help scientists understand and organize data. 
Often public outreach images are explanatory, with 
the additional requirement of piquing the public's interest. Enroute to producing outreach images the astronomer image-maker is immersed in visualizing the data, producing renditions that not only indicate which features may be significant, but versions that also generate scientific questions and discussion amongst collaborators.  Not only can this image construction guide data  analysis but it can also inspire the acquisition of additional data. A trace of this exploration is often retained in the final image, making it a synergistic hybrid of the explanatory and exploratory categories. 

\begin{figure}[h]
\centerline{\psfig{file=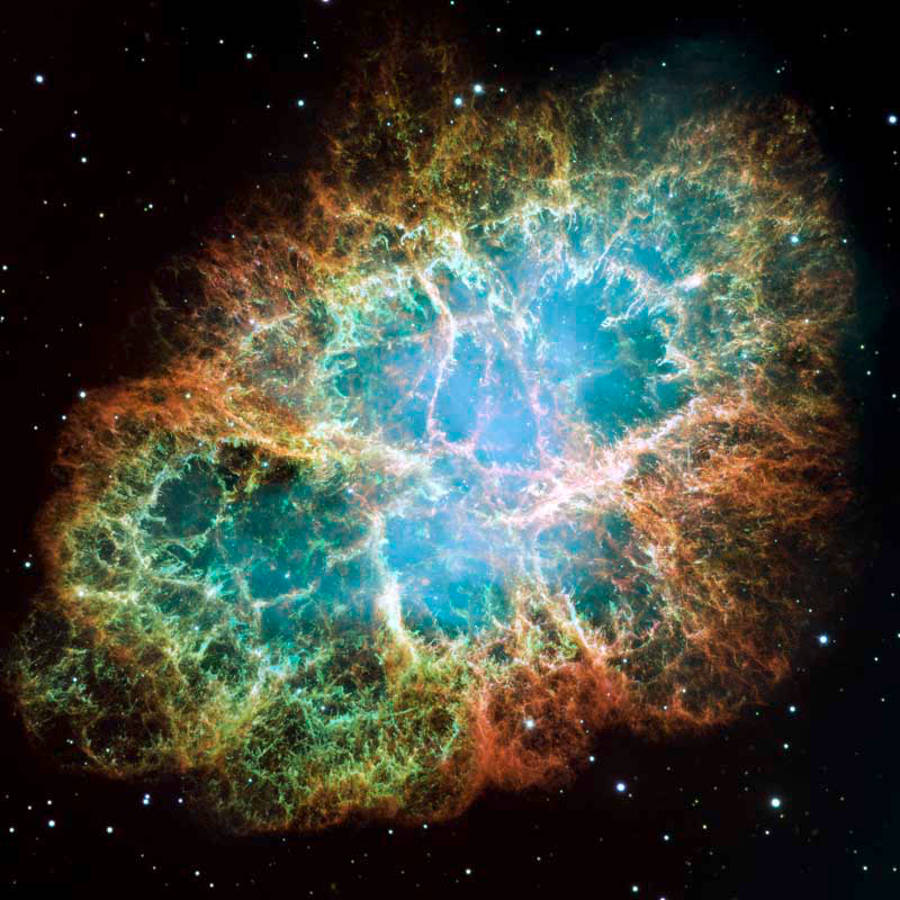,width=2.95in}} 
\caption{The Crab Nebula by Loll and Hester. 
This is an example of a public outreach image (2005) 
that has also been used in a scientific journal (2008)\cite{crab2005}. The data were acquired 
using HST's WFPC2 detector 
and 3 narrow band filters, with central transmission wavelengths at 
502, 631, and 673 nanometers. These transmit the emission from [O III],  [O I], and [S II], which have been assigned red, blue, and green respectively. In spite of this composite ordering of colour, this image is convincing in the image tradition sense. Rather than reorienting to guide the eye, this composition retains the convention of north to the top and east to the left. Comparing this to Figs.\/ \ref{figMalEngCrab} \& \ref{figCrabBlairHST} emphasizes that more than one rendition of colour and composition, using similar data of a particular target, can be successful in both aesthetics and exposition of scientific content.}
 \label{figHesterCrab}
\end{figure}

Possibly both the images' utilitarian  explanatory function and the scientists' attachment to the exploratory process led to such outreach images appearing, unexpectedly, in professional research publications. A couple of the early examples for the Hubble Heritage Project are the Elmegreen et al.~2000\cite{Elmegreen2000} paper noted above and the image of NGC~4650A in Gallagher et al.~2002\cite{Gallagher2002}.  Fig.~\ref{fighcg31final} exists because Jane Charlton arranged that  image tradition renditions of data were integral to the suite of papers\cite{hcgPalma,hcgGallagher,hcgKonst} produced by her team investigating Hickson Compact Groups. 
Completing this paper's Crab Nebula thread, J.~Hester used his team's public outreach image, Fig.~\ref{figHesterCrab}, as a figure in his review article on this ``astrophysical chimera"\cite{crab2005} .  

\begin{figure}[h]
\centerline{\psfig{file=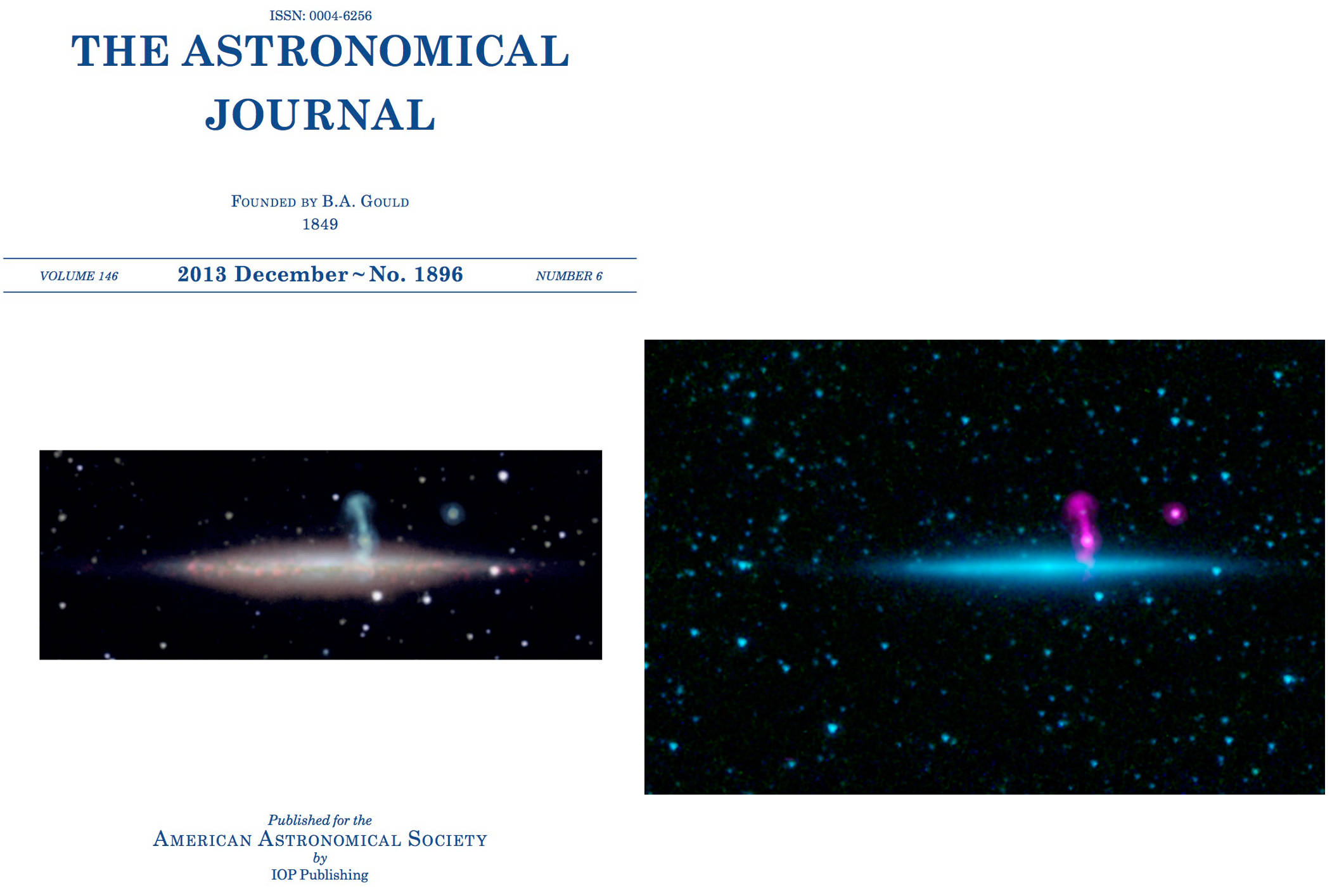,width=5.0in}} 
\caption{Comparing Public Outreach Images of UGC 10288.  Left: 
My image\cite{ugc10288} was released on National Radio Astronomical Observatory and NASA's Jet Propulsion Laboratory websites.  Galaxy UGC 10288, in warm colours, contains data from Kitt Peak National Observatory 
 Sloan Digitized Sky Survey,  
Spitzer, JVLA, and Wide-field Infrared Survey Explorer telescopes. The cooler coloured AGN jet is observed using the JVLA.  This image shows the discovery that the jet is not associated with the foreground 
galaxy but is in the distant background.   Right: The image by Robert Hurt, released on NASA's Spitzer Space Telescope website, only uses Spitzer data for the foreground galaxy. While it distinguishes between the galaxy disk and the jet, using contrast of hues (\S~\ref{contrasts}), it emphasizes the jet by using a warm colour that brings it forward.}
 \label{figugc10288}
\end{figure}

It is expected that a map produced to simultaneously be an outreach image and journal figure\cite{ugc10288} may be rather contour-like.  Ironically those produced for the public can be even more so if the  focus is on being explanatory --- see Fig~\ref{figugc10288}.

The velocity field in Fig.~\ref{figvelfield} is an example of incorporating visualization experience into explanatory research figures to support the information gleaned from the data. I would like to see colour theory (which is really colour ``experiment'') incorporated into new exploratory analysis tools as well as into standard colour tables in existing analysis software packages. Unfortunately visual literacy considerations continue to have low priority on software specification lists.  

Colour plays an important role in my own software experiments.   
I have used ``processing''\footnote{https://processing.org}, a sketchbook and a language for learning how to code within the context of the visual arts,  to create interactive tools for comparing models and/or data.  For example,  2 models can be assigned complementary colours. In locations where the models are equivalent the resultant colour is white --- otherwise their spatial distributions are two pure distinct hues.  Stepping though the channel maps interactively allows a detailed comparison.  
This tool has allowed us to visually rank models which would otherwise be judged as equivalent given their global $\chi^2$ values. 

Computational astrophysicist Gilles Ferrand and I have initiated an additional, and challenging, visualization experiment  in conjunction with Pourang Irani and his Human Computer Interaction Lab at the University of Manitoba. The lab has an immersive 3D virtual reality system that includes stereo vision with head-tracking --- our 3D spectral line cubes of galaxies' HI disks can be suspended in the 3D lab space such that the viewer can walk both around and through the data.  This situation may provide the most natural way of interacting with complex data. Even with the application of automation to ``big data'' (e.\/g.~from the upcoming Square Kilometre Array radio telescope) exploratory visualization will still be critical to discovery.  Thus a number of astrophysicists, such as Chris Fluke and David Barnes\cite{FlukeBarnesDisplay}  along with  Amr Hassan, are developing 3D immersive environments. To facilitate both the exploratory and explanatory aspects  our group would like the assignment of colour to data to be dynamic and interactive in real time. While the outcomes are not clear with regard to the efficacy of immersive tools, we hope to demonstrate the validity and effectiveness of using the principles of visual literacy in this environment.   

\section{Future Engagement with the Public: Science in Culture.}\label{future}
The evolution of practices in astronomy 
impact our engagement with the public (\S~\ref{futureastro}). Additionally science has become relevant to practices in the arts (\S~\ref{culture}).

\subsection{Feedback from Evolving Astronomy practices into Public Outreach Visuals}\label{futureastro}

New ``big science'' observatories are influencing the content and form of outreach images.  These facilities, all with international participants,  are committed to providing final data products, some even without a proprietary period for the scientists who won the competitions for observing time. They include the upgraded Jansky VLA (JVLA). 
A new low frequency regime has opened up with LOFAR (Netherlands) and international Murchison Widefield Array (Australia) observatories.  Most impressively the new Atacama Large Millimetre/submillimetre Array (ALMA; Chile) can achieve resolution better than HST and has established the microwave regime as particularly important for serendipitous discovery and the generation of conclusive results. Radio astronomy is poised to take a huge leap with the Square Kilometre Array (SKA)  and its precursor telescopes, MeerKat and the Australian Square Kilometre Array Pathfinder (ASKAP). 
Optical astronomy will benefit from the new class telescopes with  mirrors 30 m in diameter and the impending 2018 launch of the James Webb Space Telescope.  Of course a whole new window on the universe has opened up with LIGO's detection of gravitational waves and will continue to be explored with the space-based Laser Interferometer Space Antenna (LISA). 

Additionally the practice of astronomers dedicating their careers to one instrument in one regime of the electromagnetic spectrum has given way to ``multi-wavelength astrophysics".   That is, in order to answer a particular scientific question astronomers use data from the  number of radiation regimes that are relevant. Also they rarely travel to observatories to operate telescopes. (It is a shame that we can't go into space to the Hubble!) Data acquisition 
is more efficient with ``queue'' observing, which matches   
an observing program 
to the photometric conditions of that particular night. 
The requested observations are acquired by an on-site astronomer and delivered via the internet to the proposers. Other astronomers  typically have access to the data a year later.  
Processed data (step~\ref{process}, \S~\ref{construction}) can often be downloaded from observatory archives. Thereby an astronomer anywhere in the world can explore various regimes of the EM spectrum without being an expert in the esoterica of a telescope's acquisition system.

The practice of simultaneously analyzing the data from different telescopes means that figures like Fig.~\ref{figugc10288}, which incorporate data from a number of telescopes, should become more common. However,  institutions that promote single telescopes, like HST's Space Telescope Science Institute, are reluctant to produce press releases (PR) which do not emphasize HST results above all others. PR is often the first step in launching an image into the world --- the internet 
milieu described in {\S~\ref{intropower} finds its 
astronomy content via these releases. An institution (research or academic) often sends their material  to umbrella PR offices, such as those of NASA, ESA, the American Astronomical Society, and the Royal Astronomical Society, which distribute the PR more broadly and more effectively.  Hopefully the institutes producing PR will change their dissemination policies in order to reflect the fact that research astronomers wish to combine data from various sources; or perhaps  alternative, more accommodating modes of dissemination will develop. 

The use of multi-wavelength data has a direct impact of on the creation of outreach images as well.  Suppose, for example, that one wants to combine radio data and optical data. The radio data may be brightly saturated in the region where the optical observations of the target have interesting features.  To ensure the optical data information is not overwhelmed by the 
radio data requires sophisticated image manipulation techniques such as masking.   This has been employed for some time in displaying nebulae data from a single telescope, such as the Orion Nebula by David Malin\cite{malintech} and the Eskimo Planetary Nebula by Zolt Levay\cite{travis2007}.  The right-hand image in Fig.~\ref{figradiohalo} provides an example of masking JVLA data so that the dataset can be combined with HST data in order to create an image that feels more organic and less like the cut-and-paste montages that are often currently generated.

\begin{figure}[h]
\centerline{\psfig{file=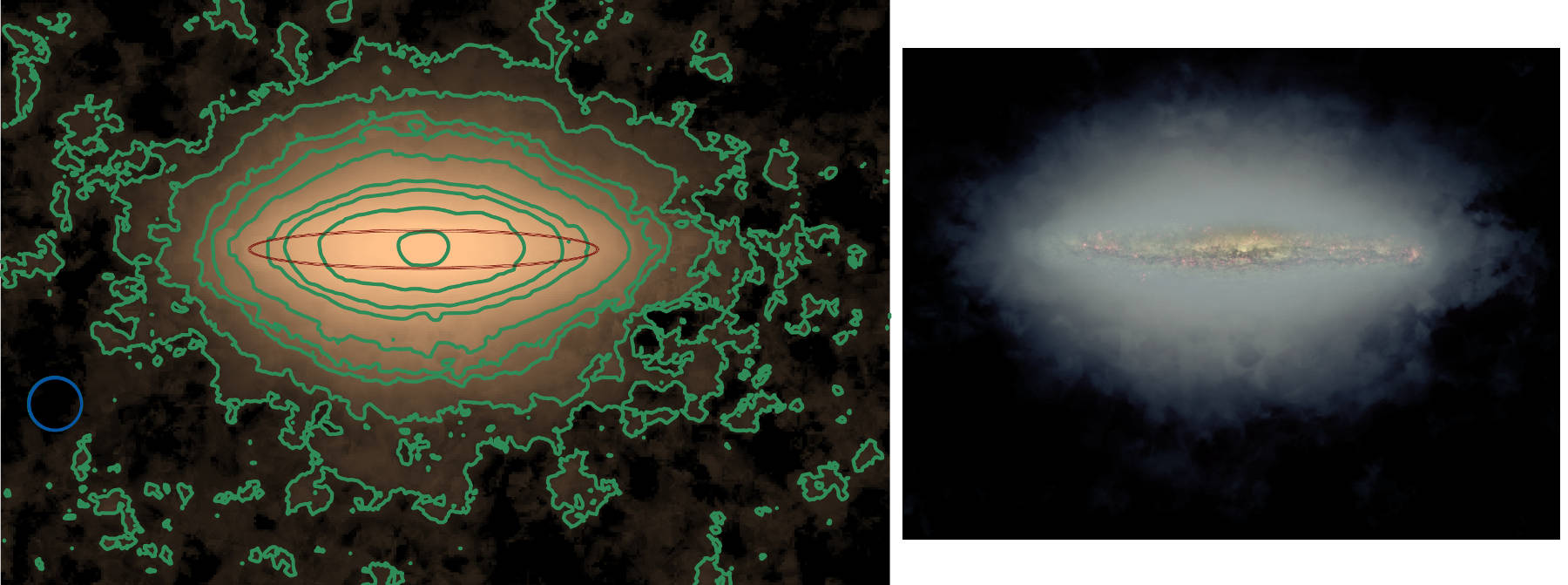,width=5in}} 
\caption{Two versions of a ``stack'' of radio halos. (\S~\ref{futureastro}.)  Left: Contour plot of the median radio halo, overlaid on its monochrome image of intensities,  published in a research journal\cite{WiegertCHANGES4}.   
The outermost contours trace noise in the data. The red ellipse delineates the area of the star forming disk. 
The blue circle provides the size of the JVLA's resolution element (synthesized beam).   Right: The version created for press releases. 
The same monochrome radio data 
 is used and my HST image of NGC 5775 is substituted for the ellipse. The halo is the same size as on the left; however, low signal-to-noise emission levels between the outer and next contour are de-emphasized.}
 \label{figradiohalo}
\end{figure}

The exquisite data from 
contemporary observatories are also evolving our modes of data analysis. For decades astronomers have  
aligned and combined different  exposures of a single resolved target through a single filter and have ``stacked'' together spectra from different unresolved
targets of the same phenomenological type. (For resolution, see \S`\ref{construction}.)   Both these approaches increase the signal-to-noise ratio in the analyzed 
data.  Currently astronomers are exploring the ``stacking'' of  resolved individual targets, as  in Fig.~\ref{figradiohalo}.  The radio synchrotron halo of 30 edge-on galaxies were scaled to a common physical size, 
where size was estimated from the extent of an ellipse encompassing the area of star formation in each galaxy. They were then aligned, to form a stack, and medianed together.  That is, at each pixel position the median intensity value was calculated for the stack of data.   This generated a single 2D dataset of median values that represents a typical radio halo. While the paper's  referee and its authors were pleased with the plot of this halo (left in Fig.~\ref{figradiohalo}) 
the institute launching the PR 
does not use contour plots. Thus an HST image,  created as described in \S~\ref{construction}, of one 
the sample's galaxies was scaled to the star forming disk 
 ellipse and masked  into a coloured and stretched version of the median radio halo dataset. ESA's HST PR office was pleased to put this image (right in Fig.~\ref{figradiohalo}) on its Facebook page.  However,  other PR offices felt the image was too much like a ``simulation'' or illustration to be promoted on their websites. 

Those PR offices have a reason to feel uncomfortable with extending astronomy analysis  practices into the public realm where they can be misread.  Western ``image tradition'' representations made of  simulations {\it generated by computers} provide an example of the issues raised.  While these simulations are not  animations of artist-made drawings, they also are not observational data.  They are visual cultural artifacts created by implementing only selected physical laws and assumptions.  They are illustrations that articulate a particular view of the science that might be occurring in a certain situation. For a scientist they are fantastically illuminating and intuition inducing.  However,  the images generated by simulations are so incredibly compelling and convincing that the public believe, for example, that LIGO has {\it literally photographed} the merger of 2 black holes.   The danger is that laypeople simultaneously feel proud of the human accomplishment and believe no more of the taxpayer's money need be spent in this research direction.   Therefore we should embed in the image a qualifying statement, such as ``based on computer simulation", analogous to 
``artist's impression" which is used for ``image tradition'' illustrations. 

\subsection{Feeding into Culture}\label{culture}
Although skilled craftsmanship is evident in the production of outreach images (\S\ref{construction}),  I do not think  of these striking images as artworks. (I use ``art'' as it is generally defined by artists and not a term to lend an object extra financial value or prestige.)  Yet there is a temptation to treat striking astronomy outreach images as art, as does journalist Jonathon Jones\cite{JJonesArt}.    
Some images almost unavoidably refer to familiar pieces of painting, photography, design, cinema, etc.~since there are numerous public visuals influencing image-makers. As well, a number of cultural theorists have been studying astronomy images, 
particularly in terms of who and what have agency, which is the ability to act on the data. It is not that there are many actors (such as the workflow steps (\S~\ref{construction}) and image-makers (\S~\ref{history})) impacting the one resultant outreach image that remove it from the art realm.  Rather it is that the astronomy image-maker is not in discourse with the concerns of contemporary photography or painting, as noted by  art historian and critic James Elkins\cite{Elkins}.

Significantly, artists have become intrigued with science and those that combine art and science increase the number of ambassadors for public outreach.  Recognizing this benefit, numerous astronomy institutions have offered artist-in-residence programs (e.\/g.~University College, London's Astronomy Artist in Residence; The Search for Extraterrestrial Intelligence Institute's SETI AIR; Canada France Hawaii Telescope Art Residency,  European Space Agency/Ars Electronica's art\&science@ESA residency). The intention is not to harness these artists as messengers for delivering specified scientific propaganda into the public realm. Rather the creative works reflect the interests and cultural concerns of the artists.  For example, the award-wining, diverse SETI AIR artists\footnote{http://www.seti.org/artist-in-residence} explore everything from psychoanalysis and philosophy to optics, algorithms and story-telling. The creations of program director, Charles Lindsay, include salvaged aerospace and bio-tech equipment.  Hence their art pieces provide examples of how art can differ from `striking' astronomy images.

In Fig.\ref{fignewmedia} I offer two examples, that incorporate public outreach images, which result from a scientist collaborating with artists. 
For one, ``Colliding Galaxies: Colours and Tones'', I produced a series of videos deconstructing the public outreach image in Fig.~\ref{fighcg31final}.  Award-winning New Music composer Nicole Lizee modified and integrated these into a composition incorporating electronic music and live performance. Both the concert and the online video\cite{collide}  increased exposure of astronomy to an atypical audience.   The other piece, ``Seeing is Believing: Experiment 1", with  artist Brad Miller (University of New South Wales) provides an example of New Media. In our  online Shockwave interactive\cite{seeing} the viewer moves the mouse, attempting to align outreach images into particular astronomy categories. 

\begin{figure}[h]
\centerline{\psfig{file=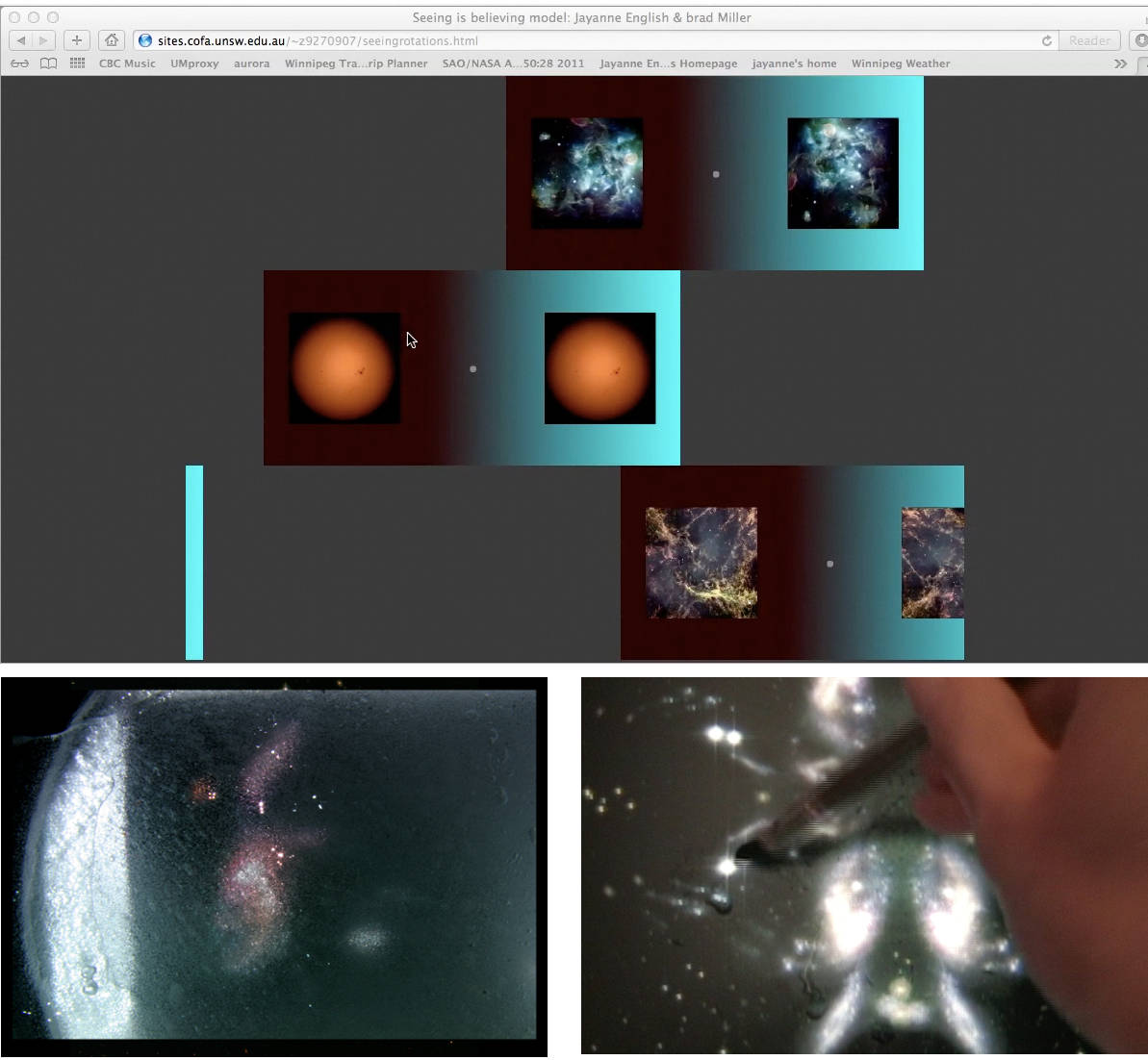,width=3.50in}} 
\caption{New Media and New Music. Top:  A screenshot of ``Seeing is Believing: Experiment 1"\cite{seeing}, a web-based interactive New Media art piece (J.~Engish and B.~Miller). As the viewer attempts to align public outreach images into particular astronomy categories fresh compositions are generated.  Bottom: Screenshots from ``Colliding Galaxies: Colours and Tones''\cite{collide} , a New Music piece that integrates sound, video and live performance (N.~Lizee and J.~English). Bottom Left:  The video features my photograph of a projection of HCG 31 (Fig.\ref{fighcg31final}) onto an ice covered mirror.  Bottom Right: Composer Nicole Lizee paints on a monitor displaying a mosaic I constructed using Fig.~\ref{fighcg31final}.}
 \label{fignewmedia}
\end{figure}

New Media melds together art praxis (fine art, architecture, and music) with science and technology (e.\/g.~engineering and computer science).    New Media is an established discipline with international conferences, dedicated organizations\footnote{e.\/g.~Subtle Technology (Canada); Australian Network for Art and Technology (Australia); Eyebeam, ASCI (USA);  European Art-Science-Technology Network; Arts Catalyst (UK)}, and degree programs at prestigious universities. Stephen Wilson's well-organized book ``Art + Science Now''\cite{Wilson}, which includes a chapter on artists focussing on the physical sciences, is lush with image documentation and online resources.

A branch called  ``artscience"  encourages one to use the creative approaches from both disciplines on any specific endeavour in either field. The most successful attempts create an object that can be appreciated simultaneously in the realm of science and in the realm of art. Bio-engineer David Edwards' book, `Artscience: Creativity in the Post-Google Generation''\cite{Edwards}, provides examples. He established an artscience lab in Cambridge, U\/.S.\/A. 
Others, like TED Fellow biophysicist Andrew Pelling\cite{Pelling} (Ottawa U.\/),  embed artists in their research labs. 

In my estimation the `composite photographs' like my own pure outreach endeavours or those by artist Michael Benson\cite{Benson}are not particularly strong examples of artscience.  
These are constructed using an approach like that outlined in \S~\ref{construction}, and  
are used in scientific contexts.  That is, the outreach images appear as illustrative, explanatory figures in professional journals and Benson's have been exhibited in the Smithsonian Institution National Air and Space Museum and Natural History Museum in London. While they can {\it depict} quantitative results (e.\/g.~Fig.~\ref{figcygnus}) and are marvellous in their public outreach capacity, our images do not produce or generate quantitative scientific results, which is a hallmark of artscience. 

Stronger,  and 
very public,  examples are the wormhole and  black hole  in the record-breaking movie `Interstellar', directed by Christopher Nolan.   
A close collaboration between Double Negative's chief scientist James Oliver and Kip Thorne resulted in the Double Negative Gravitational Renderer\footnote{As far back as 1979 Thorne's black hole was drawn, by J.\/-P.~Luminet,  as if it were photographed\cite{LuminetBH}.}. 
This iterative creative process produced scientifically constrained art objects  (the black hole and wormhole  projected on the cinema screen)  and generated several professional research articles published in prestigious science journals\cite{InterstellarJames}. 
Since this box office hit initiated 
millions of people into a physical understanding of black holes and wormholes, this film can also be perceived as an enormously successful outreach phenomenon.

Arthur I.~Miller, author of `Colliding Worlds: How Cutting-Edge Science is Redefining Contemporary Art''\cite{Miller2014},  holds the extreme, and difficult to support, position that art and science will soon again become one discipline. Nevertheless, his documentation and the above examples support Wilson's premise that scientific research, as well as technological innovation, is impacting aesthetics.  Astronomy outreach images contribute to this impact and provide an intersection point between the two cultures of art and science. Since artists' desire to collaborate with scientists continues to grow, they provide an important community for extending public outreach, albeit on their own terms.

\section{Summary and Conclusions}\label{conclusions}

Professional practices in the illustration of scientific results, such as contour plots, do not produce visuals of what the human eye sees.  However,  public outreach images, which  are in the form of  ``western image tradition'' representations  (\S~\ref{construction}), tend to convince the viewer that the image mimics nature. Partly this results from their production, which is rooted in a photographic film heritage, which in turn is often associated  with un-manipulated documentation.  Also contributing to the sense of minimal human intervention is that the image-makers are often uncredited. In reality the most striking astronomy images are constructed by skillfully applying techniques from visual art (\S~\ref{visart}) to ``logic tradition'' representations, i.\/e.~data.  This is done by teams whose members are trained in astronomical research, rather than by an individual nature photographer, with visual design training. 

These techniques, and a single workflow prescription for the construction of images (\S~\ref{construction}), can be applied to  scientific data  acquired from any energy range of the EM spectrum. Some teams represent a single institute's observing facilities and hence explore a more limited wavelength range; other teams consist of a collection of research colleagues who use data from a number of telescopes and thus sample a variety of energy ranges (\S~\ref{history}).   Since the segments of the  spectrum incorporated into outreach images often do not fall within the  wavelength limits that the human eye-brain system perceives, even a single team can create a number of different but valid renditions of a single target's data. 

Additionally there are many valid combinations of orientations, croppings, colour harmonies, contrast schemes, goals and intentions (\S~\ref{visart}).  The corollary is that many different combinations of these elements  of visual grammar will engage and retain the viewer's attention, create spatial depth,  and highlight salient features relevant to the scientists' discoveries. This is analogous to two artists having different approaches. A painting of the garden in Argenteuil by Pierre Auguste Renoir,  created while working along side Claude Monet,  is  as accomplished and relevant as the painting  of the same scene by Monet.   Similarly, still from a visual art perspective, there isn't one ``best practice in the field'' recipe for rendering an astronomical target. Although scientists may be inclined to seek a ``state-of-the-art" prescription for creating public outreach images (\S~\ref{litreview}), this paper shows that the phase-space to search is large and there isn't one global   
optimum. Indeed, even if one decide's to use a colour simply because it is their favourite, one can create an harmonious image that resonates with a viewer's physiological reading of visual stimuli (\S~\ref{perception}).
Therefore the astronomy image-maker  needs to experiment, observing what combinations work for a particular target as well as for their scientific message.   This  can lead  to improvements in visual communication as well as generate images that are powerful culturally and even politically (\S~\ref{intro}).

While compositions are arranged and colours are assigned with the above 
aesthetics in mind,  it is important to the image-making astronomers  to retain and highlight physical processes (\S~\ref{scimean}). Like their diagrammatic cousins, the astronomy images delineate scientific results, rather than correspond to what the human eye would see. Also these image-makers attempt to represent the science in a way that astronomy institutions and colleagues respect, since there is a significant relationship with the astronomers who collected the data.  For example, coming full circle the Hubble Heritage team has worked with their inspiration, Jeff Hester\cite{veilnebula}(2007).   These factors contribute to the incorporation of public outreach images as figures  professional journal articles. 

While the visual literacy acquired while constructing outreach images can enhance the explanatory quality of research publication figures (\S~\ref{feedback}), it would also be beneficial to apply visualization knowledge to software tools developed for the analysis and exploration of data. While scientists have yet to see the advantages of the most challenging visualization environments, such as virtual reality, the experiment of  making novel VR tools needs to be attempted. Employing visual grammar should enhance the chances of serendipitous positive outcomes in such experiments.  In the opposite direction, image-making methods are evolving to accommodate influences from ``big science" observatories and multi-wavelength astronomy. 

The images are disseminated in media 
ranging from newspapers through to web video podcasts. While the expected audiences consist of those predisposed  to science, typically by formal education, they have captured the interest of cultural theorists  and art-makers. The former (\S~\ref{litreview}) struggle with digesting image-making practises and unfortunately often rely on an early study that is inapplicable to current  and evolving (\S~\ref{futureastro}) imaging methods.  However the artists  (\S~\ref{culture}), particularly those working in the realms of New Media Art and Artscience, can  extend the power of engagement of outreach images to unexpected and receptive audiences.  
Recognizing this, a few astronomy institutes have set up artist-in-residence programs and some individual researchers interact with artists. 

This article attempts to inform physicists that, while ``seeing isn't believing'', science can be  rendered  both engagingly and rigorously using visual grammar so that it is relevant to scientists, artists and the general public.    Indeed if the attempt to balance communication of scientific information and perceptual power is described as a struggle between the culture of science and the culture of art, then in astronomy public outreach images both sides win. 

\section*{Acknowledgments}

The author thanks Shelley Page for a review of the manuscript and colleagues who reviewed 
sections: David Malin,  Lisa Frattare, Zolt Levay, Gilles Ferrand, and Robin Kingsburgh.   Thanks to Saeed  Salimpour for sharing his 
astrophotography report.

\end{document}